\documentclass{aa} 

\usepackage{graphicx}
\usepackage{amsmath}
\usepackage{txfonts}
\usepackage{natbib}
\bibpunct{(}{)}{;}{a}{}{}
\begin{document}

     \title{Studies of stationary features in jets: BL Lacertae I. The dynamics and brightness asymmetry on sub-parsec scales}
    
    \author{T.G.\ Arshakian\inst{1},
        A.B.\ Pushkarev\inst{2,3},
        M.L.\ Lister\inst{4},
        T.\ Savolainen\inst{5,6,7}
        }
    
        \institute{
        Byurakan Astrophysical Observatory, Aragatsotn prov. 378433, Armenia \\
        \email{t.arshakian@gmail.com}
                        \and
        Crimean Astrophysical Observatory, Nauchny 298409, Crimea, Russia
                        \and
        Astro Space Center of Lebedev Physical Institute, Profsoyuznaya 84/32, Moscow 117997, Russia
                        \and
        Department of Physics, Purdue University, 525 Northwestern Avenue, West Lafayette, IN 47907, USA
                        \and
        Aalto University Department of Electronics and Nanoengineering, PL 15500, FI-00076 Aalto, Finland
                        \and
        Aalto University Mets\"ahovi Radio Observatory, Mets\"ahovintie 114, FI-02540 Kylm\"al\"a, Finland
                        \and
        Max-Planck-Institut f\"ur Radioastronomie, Auf dem H\"ugel 69, D-53121 Bonn, Germany
                     }
    
    \date{Received ...; accepted ...}
    

  \abstract  
  {Monitoring of BL Lacertae
at 15 GHz with the Very Long Baseline Array (VLBA)  has revealed a quasi-stationary radio feature in the innermost part of the jet, at 0.26 mas from the radio core. Stationary features are found in many blazars, but they have rarely been explored in detail. }
   {We aim to study the kinematics, dynamics, and brightness of the quasi-stationary feature of the jet in BL Lacertae based on VLBA monitoring with submilliarcsecond resolution (subparsec-scales) over 17 years.
   }
   {We analysed position uncertainties and flux leakage effects of the innermost quasi-stationary feature and developed statistical tools to distinguish the  motions of the stationary feature and the radio core. We constructed a toy model to simulate the observed emission of the quasi-stationary component.
   }
   {We find that trajectories of the quasi-stationary component are aligned along the jet axis, which can be interpreted as evidence of the displacements of the radio core. The intrinsic motions of the core and quasi-stationary component have a commensurate contribution to the apparent motion of the stationary component. During the jet-stable state, the core shift significantly influences the apparent displacements of the stationary component, which shows orbiting motion with reversals (abbrev.). 
  }
   {Accurate positional determination, a high cadence of observations, and a proper accounting for the core shift are crucial for the measurement of the trajectories and speeds of the quasi-stationary component. Its motion is similar to the behaviour of the jet nozzle, which drags the outflow in a swinging motion and excites transverse waves of different amplitudes travelling downstream. A simple modelling of the brightness distribution shows that the configuration of twisted velocity field formed at the nozzle of the jet in combination with small jet viewing angle can account for the observed brightness asymmetry.}

   \keywords{BL Lacertae objects: individual (BL Lacertae) -- Galaxies: active -- Galaxies: jets -- Methods: data analysis -- Techniques: interferometric}

\titlerunning{Studies of stationary features in jets}
\authorrunning{T.G. Arshakian et al.}

\maketitle

\section{Introduction}
\label{sec:introduction}
Quasi-stationary radio components have been detected near the radio core of the jet in a number of blazars using high-resolution  VLBA monitoring at 15 GHz and 43 GHz \citep[][and references therein]{cohen14}. Stationary components appear downstream from the radio core at a distance of about $10^4$ to $10^8$ gravitational radii of the central black hole in a variety of blazars, such as BL Lacs, FR I/II radio galaxies, and flat-spectrum radio quasars \citep{cohen14}. 
A quasi-stationary feature observed in BL Lac (15 GHz) at 0.26 mas south from the radio core has been labelled as C7 and identified  as a recollimation shock \citep[RCS;][]{cohen14}. Recollimation shocks are also evident in super-magnetosonic plasma flow from two-dimensional MHD simulations of relativistic flows in which the jet dynamics are dominated by helical magnetic fields \citep{lind89,marscher08,meier12,Fromm16}. Component C7 (hereafter C7) is the closest to the radio core bright region and all moving radio components appear to emanate from C7 with relativistic speeds. \cite{cohen15} suggested that the trajectory of moving superluminal components is regulated by the dynamics of C7 and the jet dynamics is analogous to waves produced by a rapidly shaken whip. In this model, the relativistic movements of the stationary component C7 excite transverse patterns (Alf\'en waves) in the jet, which move downstream with superluminal speeds.

Very little is known about dynamics of quasi-stationary components in blazars. Studying the trajectory and kinematics of C7 in connection with varying Doppler beamed emission is of great importance for understanding the role of the quasi-stationary component in generating the jet dynamics in BL Lac. The challenge is that the scatter of positions of C7 on the sky covers about 0.1 mas,  which is smaller than the 15 GHz image restoring beam FWHM size by a factor 6-9. Therefore, the positional errors have to be carefully evaluated before analysing the scatter of C7.

Here, we study a quasi-stationary radio component in BL Lac and use the VLBA data at 15 GHz gathered in the MOJAVE (Monitoring Of Jets in Active galactic nuclei with VLBA Experiments) programme \citep{lister09}. This paper is organised as follows. Section \ref{sec:obs} describes the observational data. In Sections \ref{sec:pos_errors} and \ref{sec:motion}, we analyse the positional errors of C7 and its  motion statistics, the trajectory of smoothed motion. and the kinematics. Section \ref{sec:C7_flux_distr} studies the on-sky distribution of brightness asymmetry of C7 and a simple model is elaborated in Section \ref{sec:model}. Finally, we present discussions and conclusions in Sections \ref{sec:discussions} and \ref{sec:summary}, respectively. 

For the BL Lac at redshift $z = 0.0686$, the linear scale is 1.296 pc mas$^{-1}$, assuming a flat cosmology with $\Omega_m = 0.27$, $\Omega_{\Lambda}=0.73$, and $H_0 = 71$ km sec$^{-1}$ Mpc$^{-1}$.

\section{Observations: quasi-stationary component}
\label{sec:obs}

There are 121 epochs of VLBA observations of BL Lacertae between 1999.37 and 2016.06, most of which were made under the MOJAVE programme and 2-cm VLBA survey \citep{lister09,kellermann98}, with the rest taken from the VLBA archive following certain data quality assessments. 

Standard calibration techniques described in \cite{lister09} were used for data reduction and imaging.  
The visibility data were model fit using a limited number of two-dimensional circular or elliptical Gaussian components (see details in Cohen et al. 2014, 2015). 

At all but four epochs, a bright quasi-stationary component C7 is present at a distance of 0.26 mas from the radio core at the position angle of ${\rm PA \approx -168.^{\circ}4}$ \citep[Figs. 2 and 3 in ][]{cohen14}. One epoch, at which the component has relative ${\rm Dec}>-0.15$ mas and stands well apart from the main cluster of positions, is excluded. The scatter of the remaining 116 positions is about 0.1 mas in the RA$-$Dec plane (Fig.~\ref{fig:C7_pos_errors}). 

Observations of BL Lac at 43 GHz \citep{jorstad05,mutel05} showed that there are three stationary components (A0, A1, and A3 as designated by Jorstad et al. 2005) in the inner region of the jet within 0.3 mas. \cite{cohen14} assumed that A0 is the 43 GHz VLBI core and A1 and A2 are stationary components that are likely to be stationary shocks. The distance of A2 component from the core is $r=0.29$ mas and its position angle ${\rm PA \approx -166^{\circ}}$. The positions of C7 (15 GHz) and A2 (43 GHz) are in good agreement, indicating that the quasi-stationary component is real.

\section{Positional errors of quasi-stationary component}
\label{sec:pos_errors}
\begin{figure}[htbp]
\begin{center}
\includegraphics[width=6.5cm, angle=-90]{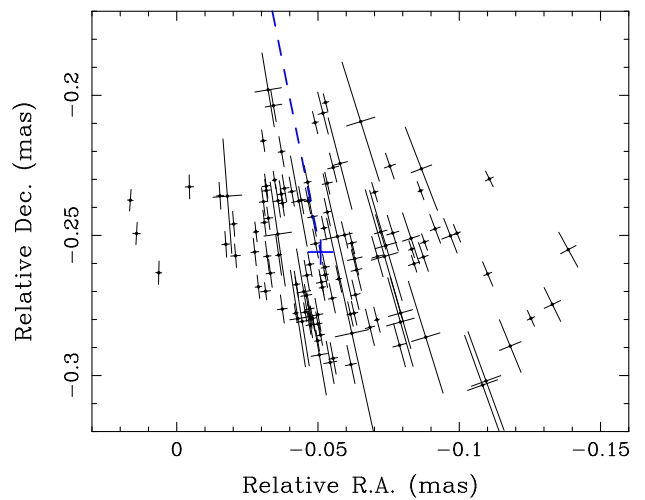}   
\caption{Distribution of 116 positions of C7 on the RA$-$Dec plane. The sizes of crosses correspond to positional errors of C7 in directions along the positional angle relative to the core and transverse to it. The median position of the scatter is marked by a plus sign. The dashed line connects the median position of C7 and radio core.}
\label{fig:C7_pos_errors}
\end{center}
\end{figure}

\begin{figure}[t]
\begin{center}
\includegraphics[width=6.6cm, angle=90]{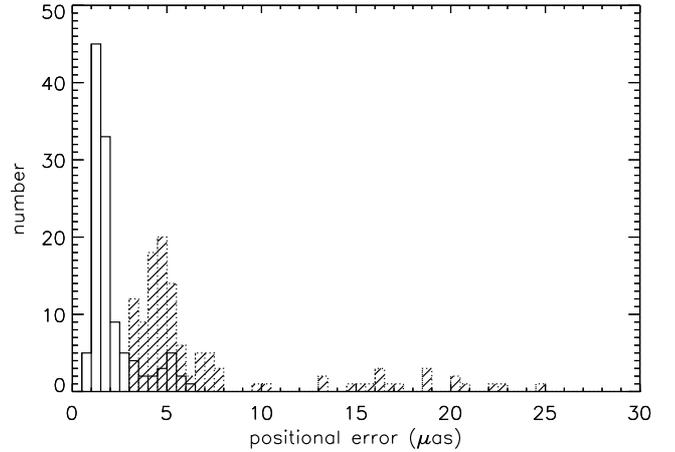} 
\caption{Distribution of 116 positional uncertainties of C7 in the direction transverse to the jet (unfilled histogram) and along the jet axis (shaded histogram). One positional uncertainty with extremely large value 0.042 mas is not visible.}
\label{fig:hist_positional_errors}
\end{center}
\end{figure}

The positions of C7, along with the corresponding positional errors between 1999.37 and 2016.06, are shown in Fig.~\ref{fig:C7_pos_errors} (dots and crosses, respectively). The positional errors are estimated using the procedure suggested by \cite{Lampton76} and based on minimising the $\chi^2$ statistic. This approach allows us to derive errors at any desired significance level $\alpha$. The position of the centroid of the C7 scatter is defined by the median values ${\rm RA_{med}}=-0.057$ mas and ${\rm Dec_{med}}=-0.254$ mas, marked by the plus sign. The radio core is located at the position RA=0 mas and Dec=0 mas outside the plot. The line connecting the radio core and the median position of C7 (dashed line) is assumed to be the jet central axis (hereafter, "jet axis"), which has a position angle of ${\rm PA}=-168^{\circ}$ (north to west). The scatter of C7 positions is quasi-circular with slight elongation along the jet central axis and has a size of $\sim 0.1$ mas (Fig.~\ref{fig:C7_pos_errors}).
\paragraph{Errors of measurements.}  The positional errors of the model-fit Gaussian components 
are estimated from the interferometric visibility plane. The fitted parameters for 
every gaussian component include its flux density, position, and size. Typically, 
the components are fitted as circular Gaussians. In rare cases, the core was 
fitted by an elliptical gaussian. As shown by \cite{Lampton76}, if $S_\mathrm{true}$ 
(``correct model'') is distributed like $\chi^2$ with $N$ degrees of freedom and the 
best-fit model, $S_\mathrm{min}$ ,with $p$ free parameters is distributed like $\chi^2$ 
with $N-p$ degrees of freedom, then the difference, $S_\mathrm{true}-S_\mathrm{min}$ , is 
distributed like $\chi^2$ with $p$ degrees of freedom. Therefore, for each data set 
at a given epoch, characterised by $N=2n_\mathrm{vis}$ degrees of freedom 
($n_\mathrm{vis}$ is the number of usable visibilities) and a model with $p$ degrees 
of freedom (a number of variable parameters minus two responsible for position), we 
change each position of the component by small, but progressively increasing, increments from 
its best fit and allow the model to relax. In {\it Difmap}, we obtain an increase in $\chi^2$,
$$
  \Delta\chi^2 = (N-p) \Delta\chi^2_\mathrm{reduced}.
$$
All three parameters, $\chi^2_\mathrm{reduced}$, $N$, and $p,$ are provided by {\it Difmap} after executing the {\it modelfit} command.  Now the limiting $\Delta\chi^2$ contour for significance $\alpha$ is
$$
  \Delta\chi^2 \leq \chi^2_\mathrm{p}(\alpha).
$$
Setting a chance probability of $\alpha = 0.32$ (68\% confidence
interval) and calculating the corresponding $\chi^2_\mathrm{p}(\alpha)$, we
find a position change that satisfies the above equation.  In this way,
we obtain ``$1\sigma$'' error. Following this procedure, we estimated
the C7 position errors in directions along the position angle of C7
relative to the core and transverse to it, also taking into account
the position uncertainty of the core itself (Fig. 1). Furthermore, for convenience, we assume that the measurement errors are measured along the jet axis and transverse to the jet axis. The C7 position
errors are highly asymmetric, with the median values $4.9$~$\mu$as and
$1.6$~$\mu$as along and transverse to the jet axis, respectively. We present the median values instead of the mean values to mitigate the effects of extreme positional errors. Distributions  of positional uncertainties are asymmetric (Fig.~\ref{fig:hist_positional_errors}) with a small secondary peak after 4 $\mu$as (unfilled histogram) and 11 $\mu$as (shaded histogram). These two peaks are interrelated and the positional errors $> 11\,\mu$as along the jet axis are mainly due to the low brightness of C7 with flux densities of less than 1 Jy. We treat the measured positional errors as lower
limits, since they assume that the complex telescope gains do not have
any errors. In fact, the gain errors may contribute to the total
budget of the position error trough phase self-calibration during
hybrid imaging. We tried to take it into account by using {\it
  selfcal} in each modelling cycle and increasing $N$ by
$(N_\mathrm{ant}-1)N_\mathrm{int}$, where $N_\mathrm{ant}$ is the
number of antennas in {\it selfcal} solution, and $N_\mathrm{int}$ is the
number of {\it selfcal} solution intervals.  The obtained error assessments,
though, show a residual inverse dependence on number of visibilities
in a data set. This indicates that the station gain phases are not
completely independent of each other, affecting the number of
parameters in a non-trivial way. Thus, the errors obtained this way
can be considered as upper limits. These are typically larger than
the lower limits by a factor of a few. Therefore, we conclude that the
true errors are somewhere between the extremes of assuming perfectly
calibrated phases and taking every self-calibrated station gain as a
free parameter. Throughout the paper, we assume that the measured positional uncertainties are close to proper errors.
\begin{figure*}[t]
\begin{center}
\includegraphics[width=6cm, angle=90]{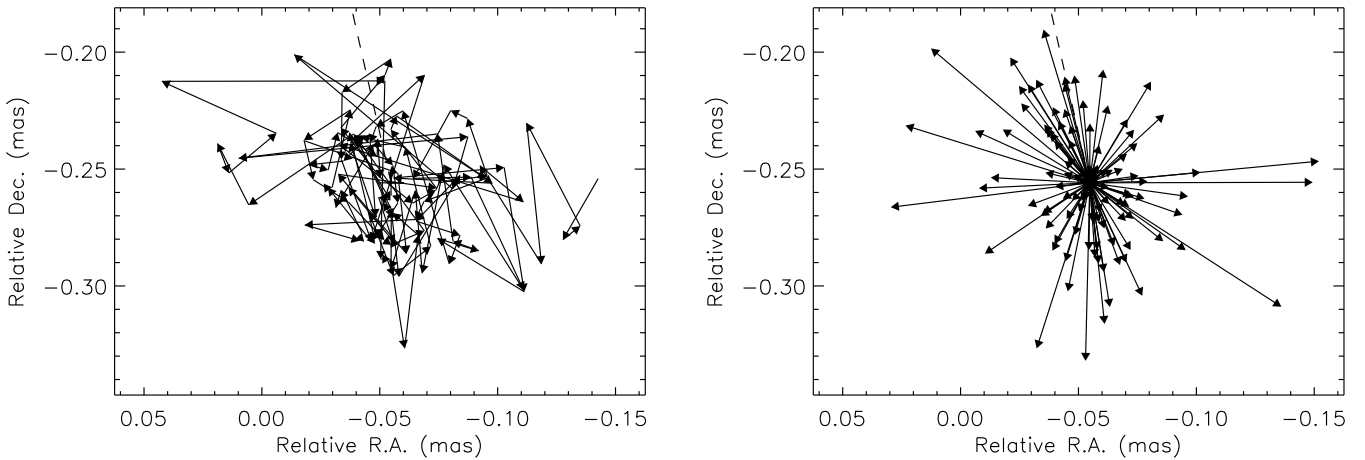} 
\caption{\emph{Left.} Track of C7 during 1999.37-2016.06. Displacement between two consecutive observations is shown by the vectors. \emph{Right.} Initial points of apparent displacement vectors in the left panel shifted to the median position of C7. The dashed line connects the median position of C7 and the radio core. The median errors of displacements along the jet axis and transverse to the jet axis are $\widetilde{\delta_{r_j}} = 5.2$ $\mu$as and $\widetilde{\delta_{r_n}} = 2.9$ $\mu$as, respectively. }
\label{fig:C7_motion_all}
\end{center}
\end{figure*}

\paragraph{Effects of flux leakage.} In order to test how much the proximity of the 
bright core component affects the apparent dependence of C7's position on its flux, we
carry out a set of simulations. For every epoch after 1999.37 with C7
present, we create a simulated data set in the following manner: 1)
Two Gaussian components representing the core and C7 are subtracted
from the actual calibrated $(u,v)$ data. The properties of these 
two Gaussians are derived from the model-fitting. 2) We then add back to
this $(u,v)$ data two Gaussians with the same positions and sizes as 
the real core and C7, but with constant flux densities of 2\,Jy and 1\,Jy,
respectively. 3) Finally, noise corresponding to the actual data
weights is added. In this way, we produce 106 visibility $(u,v)$ data
sets covering a time period of 1999.37-2016.06, and having exactly the same
$(u,v)$ coverage and noise properties as the real calibrated data. The
only difference is that now the core and C7 both have constant
flux densities. In the next step, these simulated data sets are model-fitted
by a person who did not know a priori what kind of core
structure and flux densities to expect.

As a result, the component C7 is ``found'' by the model-fit procedure in 
101 cases out of 106, with the core and C7 mean flux densities of $1.99\pm0.02$~Jy 
and $1.01\pm0.02$~Jy, respectively. No significant dependence between the C7 flux density
and its position is established. Performing these simulations, we also find that the flux leakage between 
the C7 and the core is typically small, within 10\,\%, but can reach up to 
50\,\% in rare cases.

\section{Trajectory of motion and kinematics of a quasi-stationary component}
\label{sec:motion}

\begin{figure}
\begin{center}
\includegraphics[width=6.2cm, angle=90]{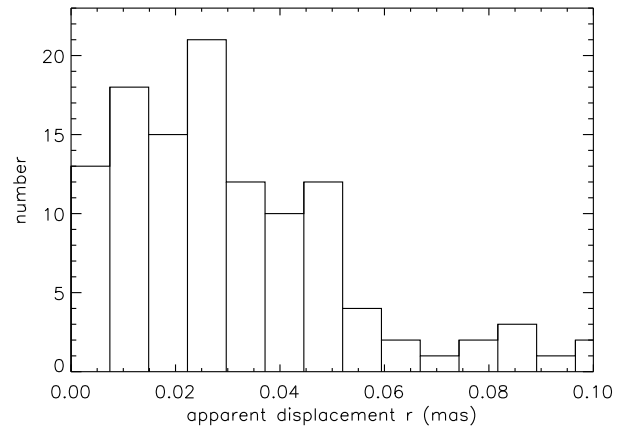}
\caption{Distribution of 115 apparent displacements of C7 between two successive epochs.}
\label{fig:hist-r}
\end{center}
\end{figure}

\paragraph{Asymmetry of displacement vectors.}
To study the motion of the C7 quasi-stationary component, we introduce the displacement vector, $\vec{r}$, which defines the direction of motion and angular displacement $r=\left|{\vec{r}}\right|$ in the sky between two consecutive epochs $t$ and $t+\Delta t$. It should be noted that the angular displacement is a combination of the cadence of observations and  motions of C7 and the core, and it rather represents the apparent angular displacements (hereafter, apparent displacements). 
The on-sky projection of trajectories of C7 between 1999.37 and 2016.06 is shown in Fig.~\ref{fig:C7_motion_all} (left panel). The orientation of the apparent displacement vectors seems random, and the majority of vectors are concentrated in a circle of diameter $\sim 0.08$ mas, which corresponds to $0.1$\,pc. To improve the visibility of orientations of the apparent displacement vectors we shift all the vectors in parallel so that their initial points coincide with median centre of the scatter of C7 positions (Fig.~\ref{fig:C7_motion_all}, right panel). There is a clear asymmetry of displacement vectors in the jet axis direction and a number of excessive long vectors are oriented in random directions. The latter account for a tail in the distribution of apparent displacements (Fig.~{\ref{fig:hist-r}}) and, therefore, six displacement vectors with the length, $r>0.08$ mas, are further excluded from statistical analysis of our sample. Those are separately examined at the end of this section. An asymmetric distribution of apparent displacement vectors can arise due to reasons of dynamics or geometry, intrinsic motion of C7, or motion of the core along the jet as a result of changes of pressure or density over time. 

\begin{figure}
\begin{center}
\includegraphics[width=6.6cm, angle=90]{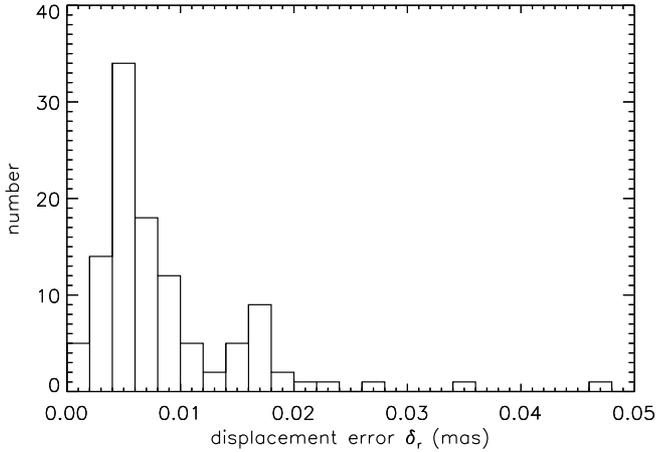} 
\caption{Distribution of 115 uncertainties of C7 displacements.}
\label{fig:hist_disp_errors}
\end{center}
\end{figure}

The errors, $\delta_r$ , associated with measured displacements are calculated by propagating the positional uncertainties of C7 at two consecutive epochs. The distribution of displacement errors is asymmetric with an extended tail towards large errors (Fig.~\ref{fig:hist_disp_errors}). The mean error of displacements is $\overline{\delta_r} = 8.5$ $\mu$as. To exclude the effects of outliers, we use the median error $\widetilde{\delta_r} = 6$ $\mu$as. We denote the projections of displacements on the jet axis and normal to the jet by $r_j$ and $r_n$ and their corresponding errors by $\delta_{r_j} = \delta_{r}\cos(\alpha)$ and $\delta_{r_n} = \delta_{r}\sin(\alpha)$, where the $\alpha$ is the angle between the displacement vector and the jet axis. The median standard errors are 
$\widetilde{\delta_{r_j}}=5.2$ $\mu$as, and $\widetilde{\delta_{r_n}} = 2.9$ $\mu$as. These errors are almost the same in all the relevant figures in the paper, unless stated otherwise. 

\begin{figure}
\begin{center}
\includegraphics[width=6.5cm, angle=90]{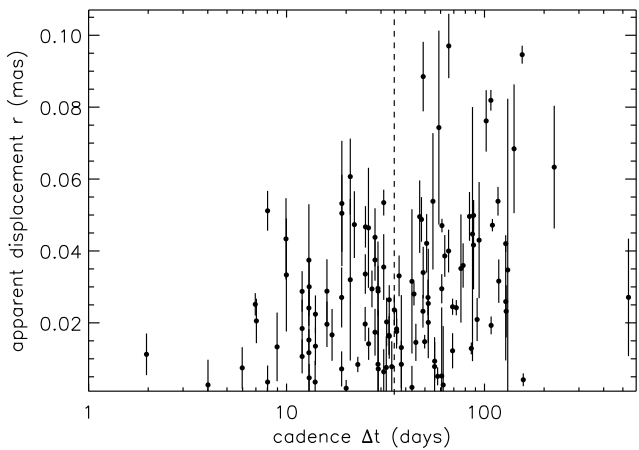}
\caption{Apparent displacement of C7 between successive epochs against observing intervals.
The$1\sigma$ error bars of the displacement ($\delta_r$) are presented. The dashed line marks a cadence of 35 days.}
\label{fig:pl-cad}
\end{center}
\end{figure}

\begin{figure}
\begin{center}
\includegraphics[width=6.1cm, angle=90]{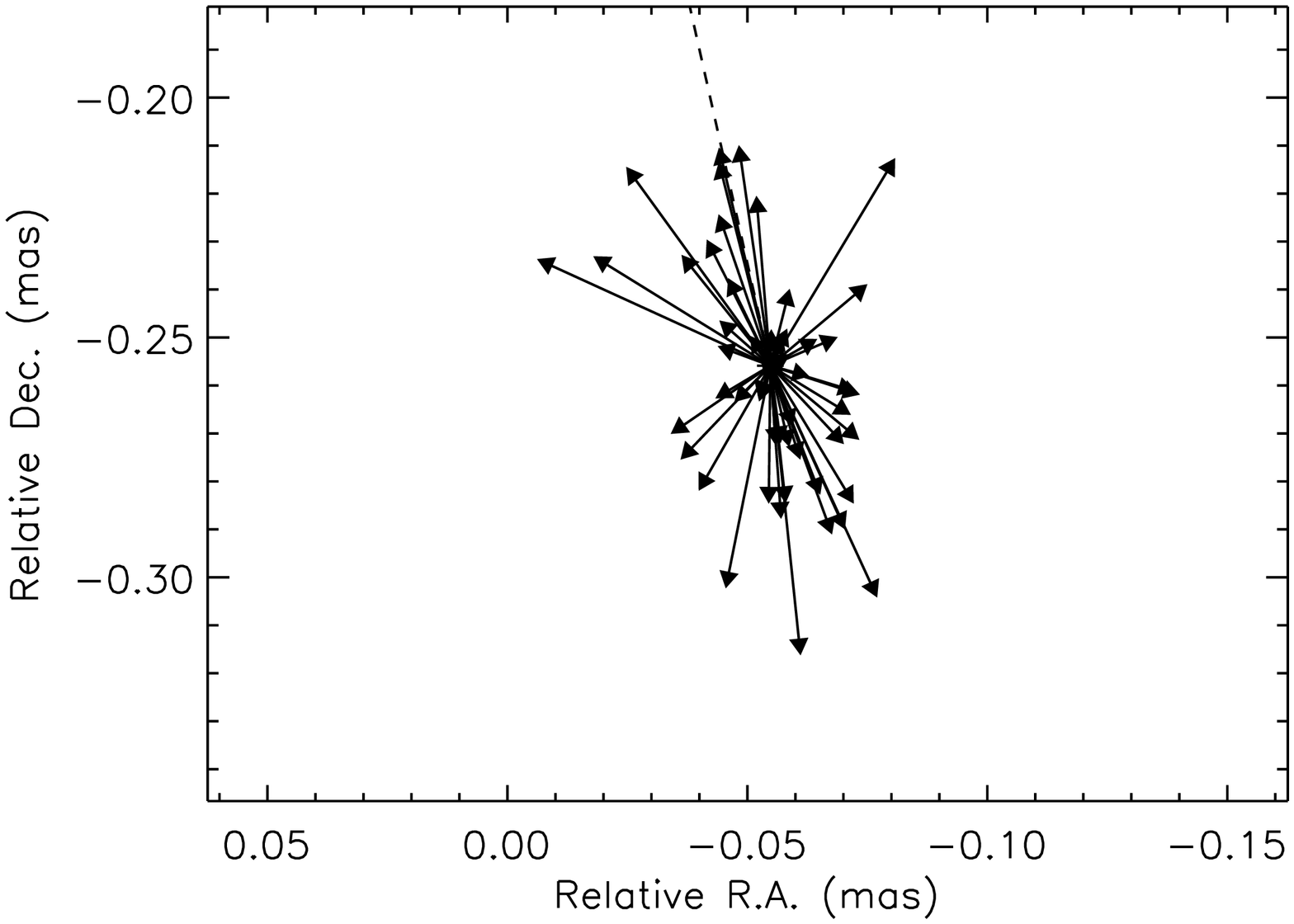} 
\includegraphics[width=6.1cm, angle=90]{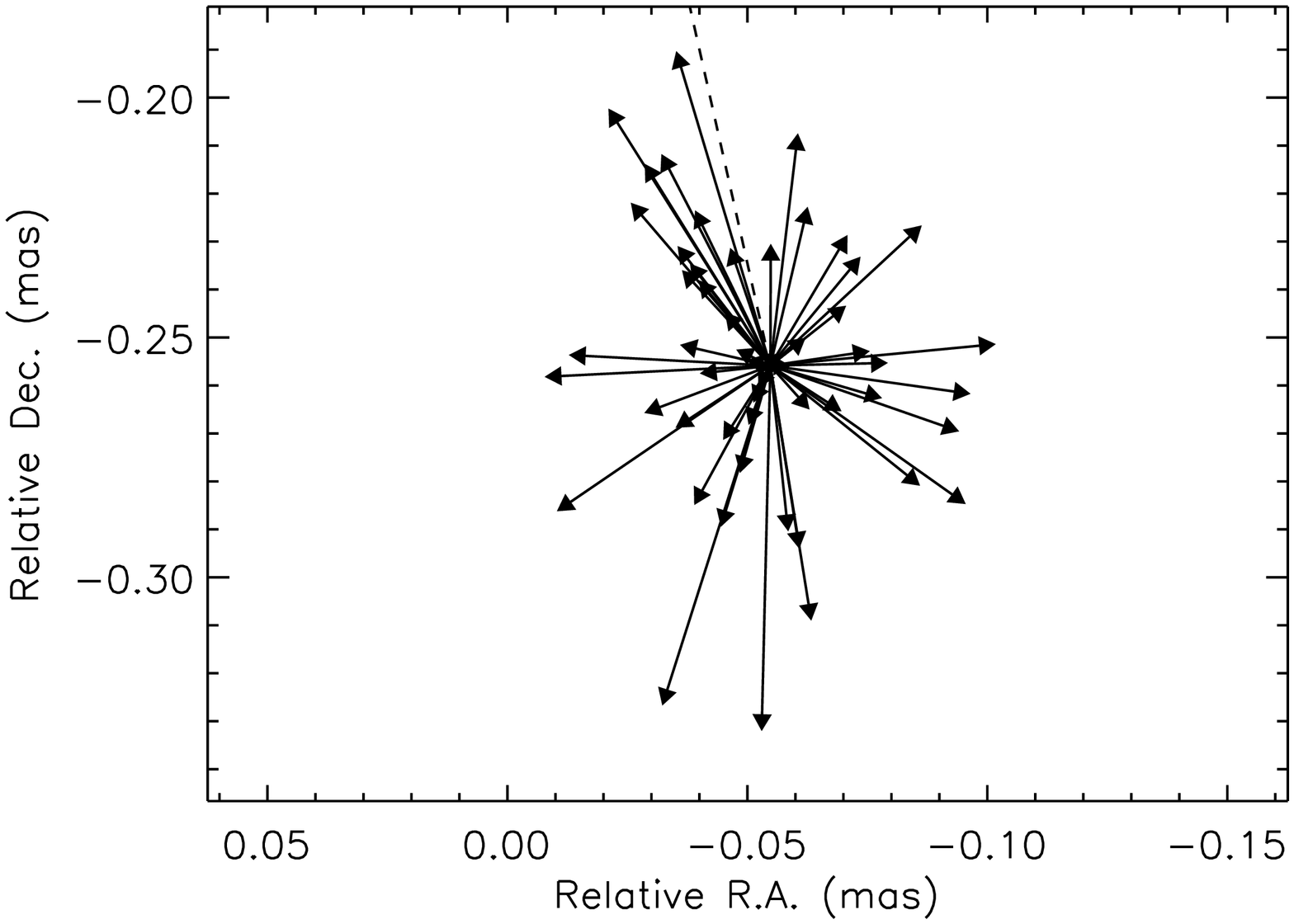} 
\caption{Apparent displacement vectors of C7 centred at the median position of C7. Trajectories observed with cadences $<35$ days (top panel) and $\ge 35$ days (bottom panel). Six apparent displacements larger than 0.08 mas are not shown. The dashed line connects the median position of C7 and the radio core.}
\label{fig:C7_motion_cadence}
\end{center}
\end{figure}

The observing intervals $\Delta t$ have a wide distribution, ranging from few days to several months (Fig.~\ref{fig:pl-cad}). If the cadence of observations is high then the intrinsic motion should be reflected in smaller observed displacements, and vice-versa. This is evident in Fig.~\ref{fig:pl-cad}, where the apparent displacements gradually increase with increasing the time $\Delta t$ between epochs. We use an interval $\Delta t=35$ days to divide the data into two comparable subsamples ($N=55$ and $N=54$). The apparent displacement vectors of C7 shifted to the median centre of the scattered positions are shown in Fig.~\ref{fig:C7_motion_cadence} for each subsample. The distribution of apparent displacements is strongly asymmetric in the jet direction for small cadences $<35$ days (top panel), while those observed with large cadences show much weaker anisotropy (bottom panel).

We examine the azimuthal distribution of apparent displacements, $r,$ to test the anisotropy of all displacement vectors. We define a polar coordinate system centred on the scattered positions and polar axis aligned with the jet direction. The azimuthal angle, $\varphi$, between the polar axis and the apparent displacement vector  
increases in the anticlockwise (eastward) direction. We rotate an angular wedge of size $\Delta\varphi=60^{\circ}$ with step of $30^{\circ}$ and calculate the number of apparent displacement vectors within the angular beam and their average  length, $\overline{r}$ , and median length, $\tilde{r}$ , for each step. 
We calculate the total variance of the mean displacement as a combination of the error variance of the mean displacement $\delta^2_{\overline{r}}$ and variance of the mean displacement, $\sigma^2_{\overline{r}}$,
\begin{equation}
        \sigma^2_t = \delta^2_{\overline{r}} +  \sigma^2_{\overline{r}} = \frac{\overline{\delta^2_r}}{N} + \sigma^2_{\overline{r}} \,,
          \label{eq:se_mr}
\end{equation}
where the first term on the right side is derived using the propagated variances of displacements and $N$ is the number of displacements. The total variance of the median displacement is also estimated using Eq.~(\ref{eq:se_mr}) but with median absolute deviation of the squared measurement error. 

\begin{figure}
\begin{center}
\includegraphics[width=6.5cm, angle=90]{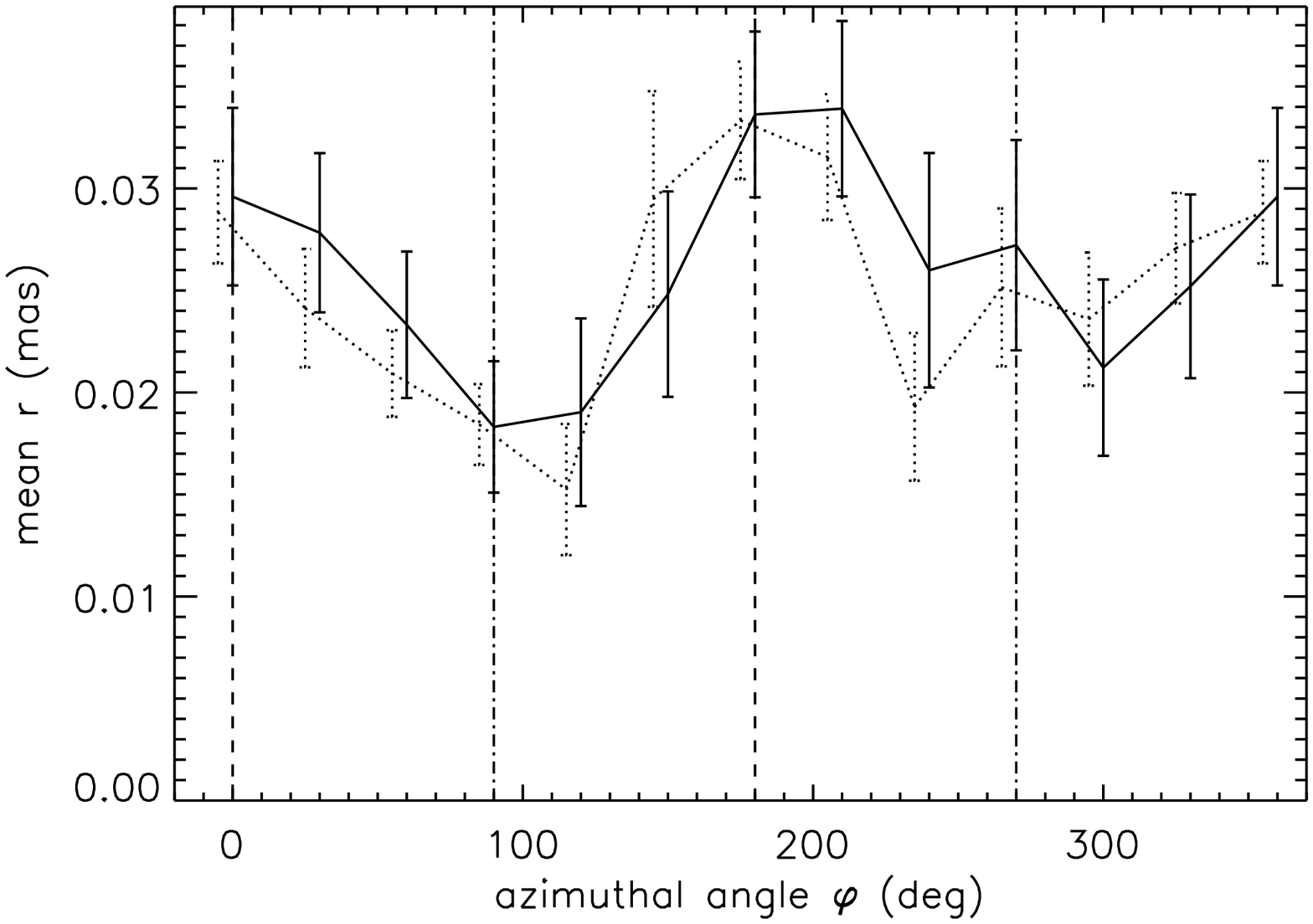} \\
\includegraphics[width=6.5cm, angle=90]{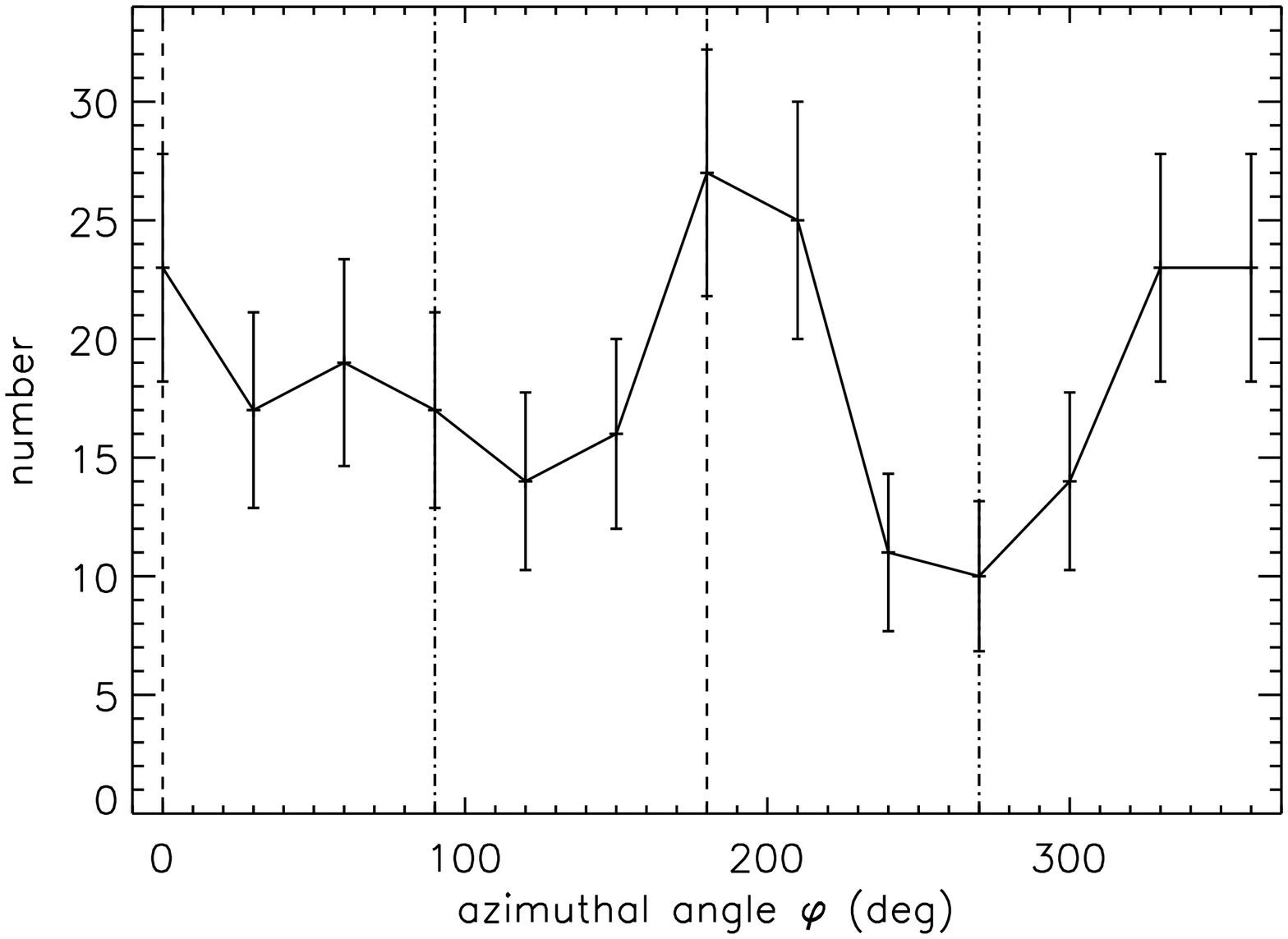}
\caption{\emph{Top.} Mean apparent displacement of C7 as a function of azimuthal angle. Mean and median apparent displacements (full line and dotted lines, respectively) are calculated for angular beam of $60^{\circ}$ with step of $30^{\circ}$. Total standard errors of the mean displacement and median displacement (Eq.~(\ref{eq:se_mr})) are presented. For illustrative purposes, the median curve is shifted to the left by $5^{\circ}$. \emph{Bottom.} Number of apparent displacement vectors of C7 against azimuthal angle. Azimuthal angle counts from the jet axis in direction downstream the jet in anti-clockwise direction. Square roots of the number counts are presented as a proxy for $1\sigma$ error. Azimuthal angles of the jet direction are marked by vertical dashed lines at $0^{\circ}$ and $180^{\circ}$, while the vertical dot-dashed lines represent the directions transverse to the jet axis, $90^{\circ}$ and $270^{\circ}$.}
\label{fig:number-azimuth}
\end{center}
\end{figure}

The mean apparent displacement varies significantly with azimuthal angle (Fig.~\ref{fig:number-azimuth}, top panel). It is longest in the direction of the jet axis at $\varphi \sim 0^{\circ}$ and $\varphi \sim 180^{\circ}$ (dashed vertical lines) and becomes shorter in directions transverse to the jet axis, $\varphi \sim 90^{\circ}$ and $\varphi \sim 300^{\circ}$ (dot-dashed vertical lines). The difference between mean apparent displacements at maxima and minima is highly significant. The distribution of is asymmetric, with a tendency to become longer in the jet direction. Surprisingly, the number of vectors varies along the azimuth in the same fashion (Fig.~\ref{fig:number-azimuth}, bottom panel), that is, it has maximum in the jet direction and reaches a minimum transverse to the jet axis. The number of apparent displacement vectors aligned with the jet $n_{\rm j}$ is about twice the number of those normal to the jet axis $n_{\rm t}$, $n_{\rm j}/n_{\rm t} \approx 2$, which is larger by a factor of two than that expected from an isotropic orientation of apparent displacement vectors. We conclude that the vector of apparent displacements of C7 are preferentially oriented along the jet and have larger magnitudes in this direction.

\paragraph{Motion of the radio core and quasi-stationary component.}
The observed anisotropy and asymmetry of the trajectories of C7 can be explained by preferential motion of either the radio core or C7 (or both) along the jet direction. The compact radio core of the jet is used in the image analysis of the VLBA observations as the reference point to measure angular displacements to stationary and moving radio features. C7 is assumed to be the nozzle of the jet that swings and generates transverse waves (Cohen et al. 2014, 2015). We can exclude the possibility that C7 swings along the jet or in the plane passing through the jet and the line of sight of the observer since the main cluster of C7 positions are distributed within $\sim 0.1$ mas around the jet, meaning that C7 moves in all directions. The shift of the core along the jet may happen due to resolution-dependent effects or changes of opacity (or in electron density) in the core region (see more detail discussions on these effects in Section \ref{sec:discussions}).

We then develop a model where the core and C7 have intrinsic motions with respect to the central black hole of BL Lac: (i) the core position wanders in the direction of the jet axis and (ii) the C7 moves in random directions within a spherical volume limited by a radius $s_{\rm max}$ of the scatter of positions. We denote the spatial displacement vectors of the C7 and core by $\vec{S}$ and $\vec{C}$ and their on-sky projections by $\vec{s}$ and $\vec{c}$, respectively, which we call the displacement vectors of the C7 and core. In the VLBA images we are able to measure only the relative (apparent) motion of C7 with respect to the core. Then the apparent displacement vector of C7 ($\vec{r}$) is the combination of the displacement vectors,  
\begin{equation}
  \vec{r} = -\vec{c} + \vec{s}.
  \label{eq:vector_r}
\end{equation}
If the  motion of the core dominates over the  motion of the C7, $c \gg s$, then $\vec{r} \approx -\vec{c}$, meaning the apparent motion of C7 is the mirror reflection of the asymmetric motion of the core along the jet axis. In the case of $c \ll s$, the core motion is negligible and we recover the  motion of C7, $\vec{r} \approx \vec{s}$. The projection of a vector on the jet axis is defined to be positive downstream the jet. The projections of the apparent displacements $r$ onto the jet axis are
\begin{equation}
  {r_j} = -c + s_j,
  \label{eq:vector_r}
\end{equation}
where $c \in [-c_{\rm max}, c_{\rm max}]$ and $s_j \in [-s_{\rm max}, s_{\rm max}]$. 

Assuming an axis transverse to the jet intersects the latter at the median position of C7. The transverse projections of the apparent displacements, $r,$ are independent of displacements of the radio core since the transverse projections of the core $c_{n}=0$, then
\begin{equation}
  r_n = s_n,
  \label{eq:vector_rn}
\end{equation}
where $s_n = s \sin(\alpha) \in [0, s_{\rm max}]$ and $\alpha \in [0,\pi]$ is the angle between the jet axis and the vector $\vec{s}$.

We wish to quantify the relative contribution of the  displacements of the radio core and C7. We designate the probability density functions of the dependent variable by $f(r \equiv |\vec{r}|)$ and independent variables by $g(c)$ and $h(s)$ and the probability density function of the spatial displacements of C7 by $H(S)$. We take the advantage of the isotropic distribution of the $S$ to relate the statistics $\overline{S}$ and $\overline{s}$ to the statistics of measurable quantities, $r_j$ and $r_n$.       
\begin{figure}
\begin{center}
\includegraphics[width=6.3cm, angle=90]{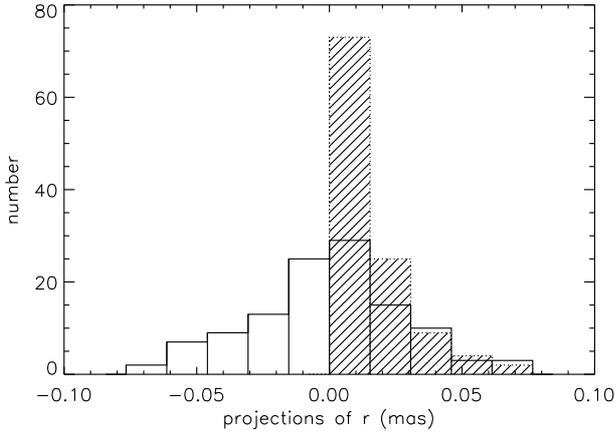}
\caption{Distributions of projections $r_j$ and $r_n$ of the apparent displacements $r$ onto the jet axis (unfilled histogram) and transverse to the jet axis (shaded histogram).  }
\label{fig:rj_rt}
\end{center}
\end{figure}

Considering that the  displacements of C7 $s = S \sin(\phi)$ (where $\phi \in [0, \pi]$ is the angle between the spatial vector, $\vec{S,}$ and the line of sight) and the vectors, $\vec{S,}$ are distributed isotropically, $K(\phi) = \frac{1}{2}\sin(\phi)$, we can derive an integral relation between the probability density functions $h(s)$ and $H(S)$ \citep[see for an example,][]{al00},
\begin{equation}
h(s) = \frac{1}{2}s \int_{s}^{s_{\rm max}} \frac{H(S)}{S \sqrt{S^2-s^2}} dS.
  \label{eq:h}
\end{equation} 
We introduce the first and second moments of the probability density function $H(S)$ by multiplying Eq.~(\ref{eq:h}) by $sds$ and $s^2ds$ and integrating from 0 to $s_{\rm max}$. We recover the mean and mean square of spatial  displacements, 
\begin{equation}
\overline{S} = \frac{4}{\pi} \overline{s} \,\,\,{\rm and} \,\,\, \overline{S^2}=\frac{3}{2} \overline{s^2},
  \label{eq:sm1}
\end{equation}
and variance of $S$,
\begin{equation}
\sigma_{S}^2= \frac{3}{2} \overline{s^2} - \frac{16}{\pi^2} (\overline{s})^2.
  \label{eq:sigma_s1}
\end{equation}
The values $\overline{s}$ and $\overline{s}^2$ are not known and our next step is meant to relate  
these quantities to a measurable parameter. We consider the projections of the  displacements $s$ transverse to the jet axis, $s_n = s \sin (\alpha)$, where $\alpha \in [0, \pi]$ is the angle between the jet axis and the vector $\vec{s}$. Using the assumption of isotropy of $\vec{s}$, $K(\alpha) = 1/\pi$, we obtain an integral relation between the probability density functions $h(s)$ and $u(s_n)$,
\begin{equation}
u(s_n) = \frac{1}{\pi} \int_{s_n}^{s_{\rm max}} \frac{h(s)}{\sqrt{s^2-s_n^2}} ds.
  \label{eq:u}
\end{equation} 
We derive the first and second moments of $h(s)$ in the same way as for Eq.~(\ref{eq:h}). Taking into account the equality, $s_n = r_n$ (Eq.~(\ref{eq:vector_rn})), we obtain the mean and variance of the  displacements of C7, which are expressed in terms of a measurable parameter $r_n$,
\begin{equation}
\overline{s} = \frac{\pi}{2} \overline{r_n} \,\,\,{\rm and} \,\,\, \overline{s^2}=2 \overline{r_n^2}
  \label{eq:means}
\end{equation}
and
\begin{equation}
  \sigma^2_{s} = \overline{s^2} - (\overline{s})^2 = 2 \overline{r_n^2} - \frac{\pi^2}{4} \overline{r_n}^2.
  \label{eq:vars}
\end{equation}

Substituting the latter into Eqs.~(\ref{eq:sm1}) and (\ref{eq:sigma_s1}), we derive the mean and variance of the spatial displacements of C7:
\begin{equation}
\overline{S} = 2\overline{r_n}
  \label{eq:mS1_rn}
\end{equation}
and 
\begin{equation}
\sigma_{S}^2= 3\overline{r_n^2} - 4(\overline{r_n})^2.
  \label{eq:sigma_S1}
\end{equation}
We use the subsample of 109 displacements with lengths of less than 0.08 mas to estimate the spatial mean displacement of C7, $\overline{S} = 0.026$ mas, and standard deviation of spatial displacements, $\sigma_{S} = 0.015$ mas, for the period 1999.37-2016.06.  

Our next task is to derive the relation between the statistics of the core and the $r_j$. 
The mean of the sum of two independent variables $c$ and $s_j$ (see Eq.~(\ref{eq:vector_r})) is
\begin{equation}
\overline{r_j} = -\overline{c} + \overline{s_j}.
  \label{eq:rjm}
\end{equation}
and the variance is given by
\begin{equation}
\sigma_{r_j}^2 = \sigma_{c}^2 + \sigma_{s_j}^2.
  \label{eq:sig_rj2}
\end{equation}
The mean, $\overline{r_j}$ , is estimated from the transverse projections $r_j$. Since the $\overline{s} = 0$ due to isotropy of vectors $\vec{s}$, then we obtain from Eq.~(\ref{eq:rjm}),
\begin{equation}
\overline{c} = -\overline{r_j}.
  \label{eq:cjm}
\end{equation}
and 
\begin{equation}
\sigma_{c}^2 = \sigma_{r_j}^2 - \sigma_{s_j}^2 = \sigma_{r_j}^2 - \sigma_{r_n}^2,
  \label{eq:sig_rj}
\end{equation}
considering that the variances of the projections of isotropic displacements,$s,$ on the jet axis and transverse to the jet axis are equal $\sigma_{s_j}^2 = \sigma_{s_n}^2$ and the equality, $\sigma_{s_n}^2 = \sigma_{r_n}^2$ (Eq.~(\ref{eq:vector_rn})). The distribution of $r_j$ is near-symmetric (unfilled boxes in histogram in Fig.~\ref{fig:rj_rt}) with mean $\overline{r_j} \approx 10^{-4}$ mas and standard deviation of $\sigma_{r_j} = 0.027$ mas, and we estimate the mean  displacement of the core $\overline{c} \approx 0$ and $\sigma_{c} = 0.025$ mas for the period 1999.37-2016.06 (Table~\ref{table1}). The same statistics of the C7 are $\overline{s} = 0.021$ mas and $\sigma_{c} = 0.014$ mas. To compare the significance of the means of the C7 and core, we use the root of the second moment for the C7 (Eq.~(\ref{eq:means})),
\begin{equation}
{\rm rms}_{s} = \left( \overline{s^2}\right)^{\frac{1}{2}} = \left( 2\overline{r_n^2} \right)^{\frac{1}{2}},
  \label{eq:rms_s}
\end{equation}
and for the core the standard deviation is the rms, $\sigma_{c} = \left ( \overline{c^2} \right )^{\frac{1}{2}}$, since $\overline{c}= 0$,
 then
\begin{equation}
{\rm rms}_{c} = \sigma_{c} = \left( \sigma_{r_j}^2 - \sigma_{r_n}^2 \right)^{\frac{1}{2}}.
  \label{eq:rms_c}
\end{equation}
The equations for estimating the errors of the rms$_s$ and rms$_c$ are given in Appendix~\ref{asec:se_rms}.  

\begin{figure*}[t]
\begin{center}
\includegraphics[width=10cm, angle=90]{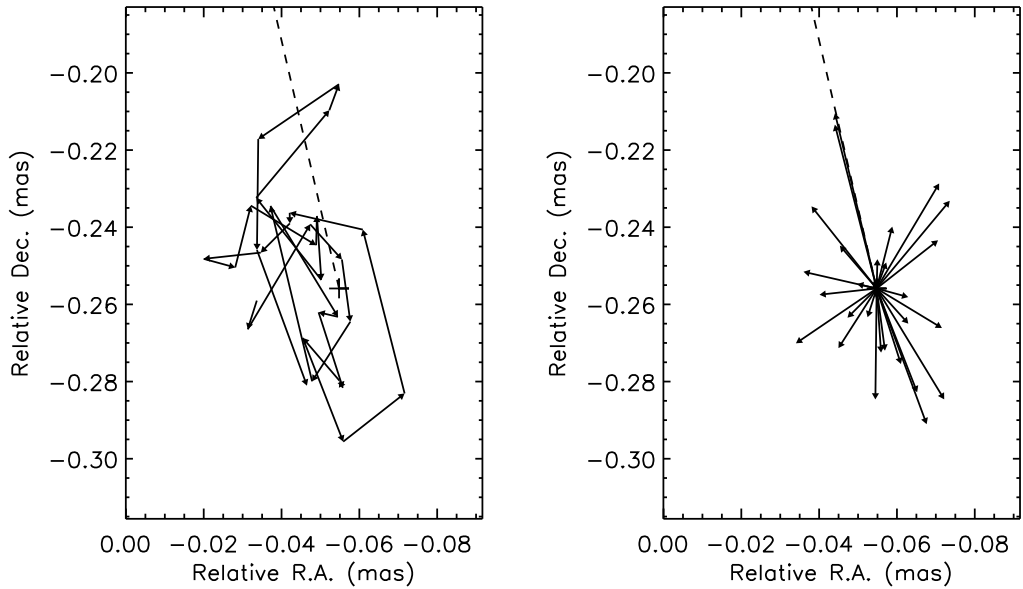}
\caption{\emph{Left.} Track of C7 component during 2010-2013. Displacement between two subsequent observations is shown by the vectors. \emph{Right.} Initial points of apparent displacement vectors shifted to the median position of C7. The dashed line connects the median position of C7 and the radio core. The median errors of displacements along the jet axis and transverse to the jet are $\widetilde{\delta_{r_j}} = 4.5$ $\mu$as and $\widetilde{\delta_{r_n}} = 2.3$ $\mu$as, respectively.}
\label{fig:C7_motion_2010-2013}
\end{center}
\end{figure*}
\begin{table*}
\caption{Estimates of the mean, standard deviation, and rms of the displacements of C7 and the core (marked in bold font). The unit of measurements is given in mas.}\label{table1}
\centering
\begin{tabular}{lccccccccc}
\hline \hline
Epoch & N & $\overline{r_n}$ & $\overline{r_n^2}$ &  $\pmb{\overline{s}}$ & $\pmb{\sigma_{s}}$ & $\pmb{{\rm rms}_s\pm\delta_{{\rm rms}_s}}$ & $\sigma_{r_j}$ & $\pmb{\overline{c}=-\overline{r_j}}$ & $\pmb{\sigma_{c}} = \pmb{{\rm rms}_c\pm\delta_{{\rm rms}_c}}$   \\
   (1)       & (2)   &    (3)              &        (4)       &             (5)                                                     &
             (6)              & (7) & (8) &      (9)              &         (10)             \\ 
\hline
$1999.37-2016.06$          &  109  & 0.013   & 0.0003   & $0.021$ & 0.014 & 0.025$\pm$0.009 & 0.027 & $\approx 10^{-4}$ & 0.025$\pm$0.008   \\
$2010-2013$          &  28    & 0.0084 & 0.00012 & $0.013$ & 0.008 & 0.015$\pm$0.005 & 0.021 & $\approx 10^{-4}$  & 0.019$\pm$0.007            \\
$\le2010$ or $\ge2013$ & 81  & 0.0152  & 0.0004  & $0.024$ & 0.014 & 0.028$\pm$0.009 & 0.029 & $\approx 10^{-4}$ & 0.026$\pm$0.010         \\                        
\hline
\hline
Cadence & N & $\overline{r_n}$ & $\overline{r_n^2}$ &  $\pmb{\overline{s}}$ & $\pmb{\sigma_{s}}$ & $\pmb{{\rm rms}_s\pm\delta_{{\rm rms}_s}}$ & $\sigma_{r_j}$ & \pmb{$\overline{c}=-\overline{r_j}$} & $\pmb{\sigma_{c}} = \pmb{{\rm rms}_c\pm\delta_{{\rm rms}_c}}$    \\
(days)    &    &                           &                               &                                                               &
                         & & &                    \\ 
\hline
$<35$          &  55  & 0.011 & 0.0002 & $0.015$ & 0.011 & 0.019$\pm$0.007 & 0.025 & $\approx -10^{-3}$ & 0.024$\pm$0.007     \\
$>35$          &  54  & 0.017 & 0.0005 & $0.027$ & 0.013 & 0.030$\pm$0.009 & 0.029 & $\approx 10^{-3}$ & 0.026$\pm$0.013     \\
\hline
\end{tabular}
\vspace{1ex}
 
     {\raggedright {\bf Notes}. Columns are as follows: (1) range of epochs and cadences, (2) number of displacements, (3) $\overline{r_n}$ is the mean of apparent displacements of C7 projected normal to the jet axis, (4) $\overline{r_n^2}$ is the mean of $r_n$ squared, (5) $\overline{s}$ is the mean of  displacements of C7 (Eq.~(\ref{eq:means})), (6) $\sigma_{s}$ is the standard deviation of the  displacements (Eq.~(\ref{eq:vars})), (7) ${\rm rms}_s$ of the displacements of C7 (Eq.~(\ref{eq:rms_s})) and its standard error (Eq.~(\ref{aeq:se_rmss})), (8) $\sigma_{r_j}$ is the standard deviation of projections of the apparent displacements of C7 on the jet axis, (9) $\overline{c}$ is the mean  displacement of the core (Eq.~(\ref{eq:cjm})),
(10) ${\rm rms}_c$ or $\sigma_{c}$ of the  displacements of the core (Eq.~(\ref{eq:rms_c})) and their standard error (Eq.~(\ref{aeq:se_rmsc})).  \par} 
\end{table*}

For the period 1999.37-2016.06, the rms of the displacements of C7 and the core have an equal contribution to the apparent motion of C7 ($\approx$ 0.025 mas, see Table~\ref{table1}).
\cite{cohen15} noticed a different jet behaviour in 2010-2013, that is, the PA of the jet varies in a small range around $-170^{\circ}.5$ and most of the jet ridge lines show weak quasi-standing wiggles. They suggest that the intense swinging of C7 in PA excites large transverse waves downstream in the jet, while the small variation of the PA generates a weak variable wiggle seen in 2010-2013. For the latter period, we plot the apparent displacement vectors of C7 (left panel in Fig.~\ref{fig:C7_motion_2010-2013}) and these vectors are centred at the median position of C7 (right panel). The apparent displacement vectors have a preferential orientation along the jet axis (dashed line). A contribution to the asymmetry of the apparent displacements of C7 is due to displacements of the core, for which the ${\rm rms}_{c} = 0.019$ mas is larger than the ${\rm rms}_{s} = 0.015$ mas of C7 (Table~\ref{table1}) but they are comparable within error limits. The contribution of C7 (the isotropic component) and the core (anisotropic component) increases during the time period beyond 2010-2013 when the jet swinging activity is high. A comparison of statistics between 2010-2013 and beyond (1999.37-2010 and 2013-2016.06) shows that during the latter periods, the ${\rm rms}$ of C7 becomes larger by a factor of about 2, while the ${\rm rms}$ of the core changes by a factor of about 1.4 in a small range between 0.019 mas and 0.026 mas. 
During the rise of the core shift activity the swinging amplitude of C7 becomes twice as large. It is likely that the motions of the core and C7 are related events. 

The smaller the observing intervals, the more realistic the trajectories of C7. We suppose that the true estimates of statistical characteristics of displacements are found for the observing intervals $\Delta t < 35$ days. The ${\rm rms}$ of the displacements of the core dominates over the rms of C7 ($0.024$ mas $>0.019$ mas), but they are indistinguishable within $1\sigma$ error (see Table~\ref{table1}), that is, the contribution of both to the apparent motion of C7 are of comparable importance. The high positional accuracy of the core and C7 and high-cadence VLBA observations at 15 GHz with an interval of less than or about a month are required to identify the dominance between the core and C7 and to study their dynamics.

We conclude that the core shift can account for both asymmetric and anisotropic distributions of the apparent displacements of C7 (Fig.~\ref{fig:number-azimuth}) and the contribution of the core shift and C7 to the apparent motion of C7 are significant during both the jet-stable state, when the swinging activity is relaxed, and the high state of swinging activity.
\begin{figure}[t]
\begin{center}
\includegraphics[width=6.2cm, angle=90]{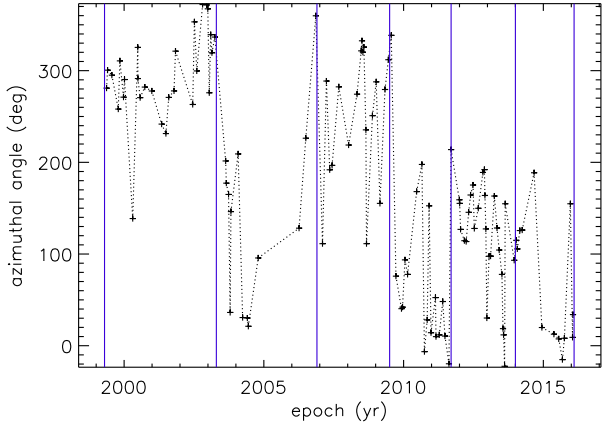}
\caption{Azimuthal angle of C7 against epoch. The vertical lines show the selected six time intervals.}
\label{fig:C7_azimuth-epoch}
\end{center}
\end{figure}

We introduce an rms for the spatial displacements of C7 and core. The rms displacement of C7 is the square root of the second moment (the second term in the Eq.~(\ref{eq:sigma_S1})), ${\rm rms}_S = \left ( 3\overline{r_n^2} \right )^{1/2}$, and rms of the core is simply, ${\rm rms}_C = {\rm rms}_c / \sin(\theta) \approx 7 \,{\rm rms}_c$, where the jet viewing angle $\theta = 8^{\circ}$. The rms of the spatial  displacements of the core between 1999.37-2016.06 is ${\rm rms}_C = 0.17\pm0.06$ mas and it is about four times greater than the ${\rm rms}_S = 0.04\pm0.01$ mas (errors are estimated from Eqs.~(\ref{aeq:se_rms_S}) and (\ref{aeq:se_rms_C}) ). In fact, the spatial statistical characteristics of the core are much higher than those of the C7, but they are strongly reduced due to foreshortening by a factor of about 7.

We estimated the statistics from Table~\ref{table1} but using the median displacement and median absolute deviation. The median displacements, median absolute deviations, and median rms are found to be smaller by a factor of $1.7\pm0.03$ than those in Table~\ref{table1}. We believe that the true values are somewhere between the mean and median values. Estimates of the proper positional errors of C7 are needed to accurately assess the statistical characteristics of the displacements.
\begin{figure}
\begin{center}
\includegraphics[width=3.5cm, angle=90]{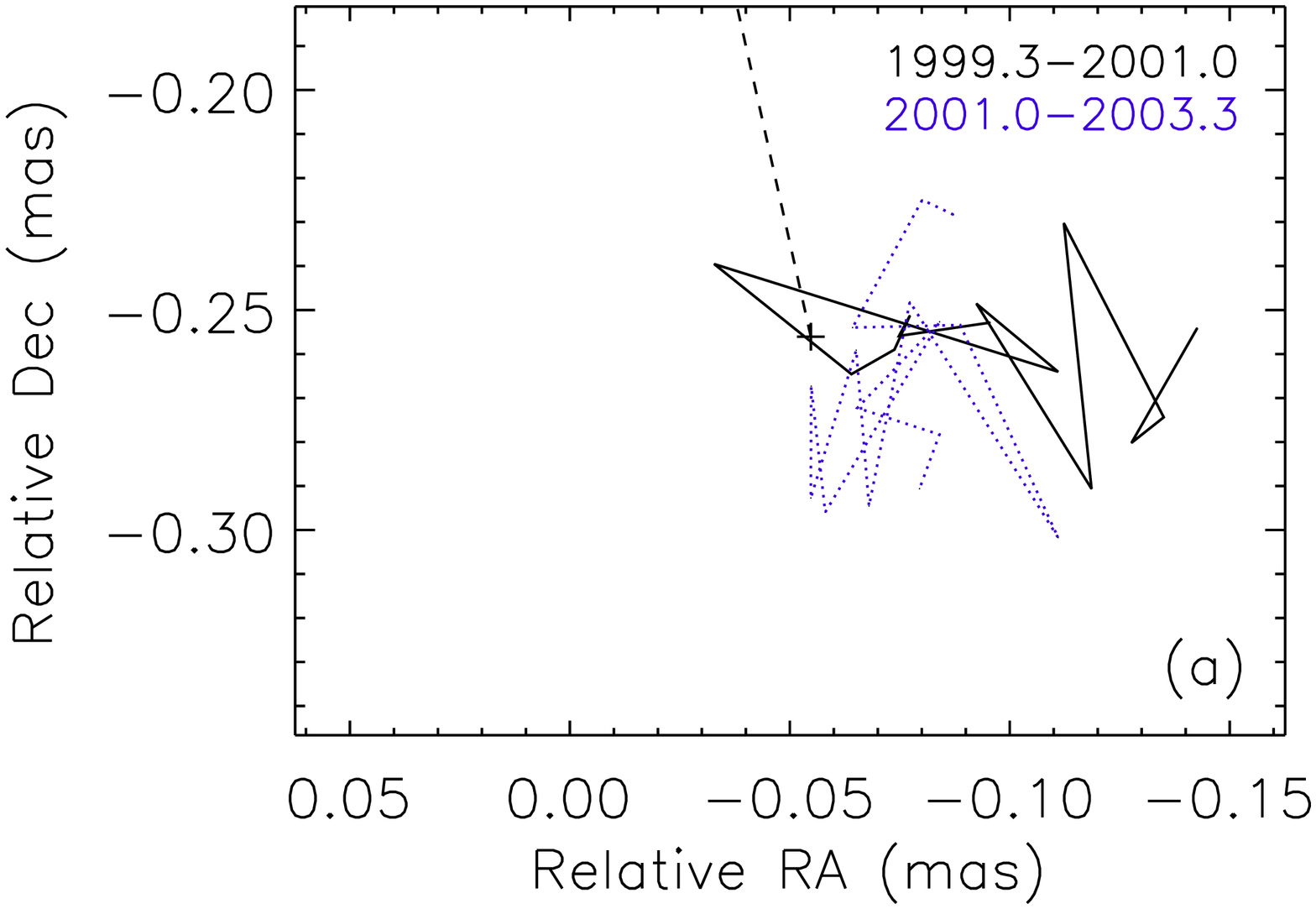}  \\
\includegraphics[width=3.5cm, angle=90]{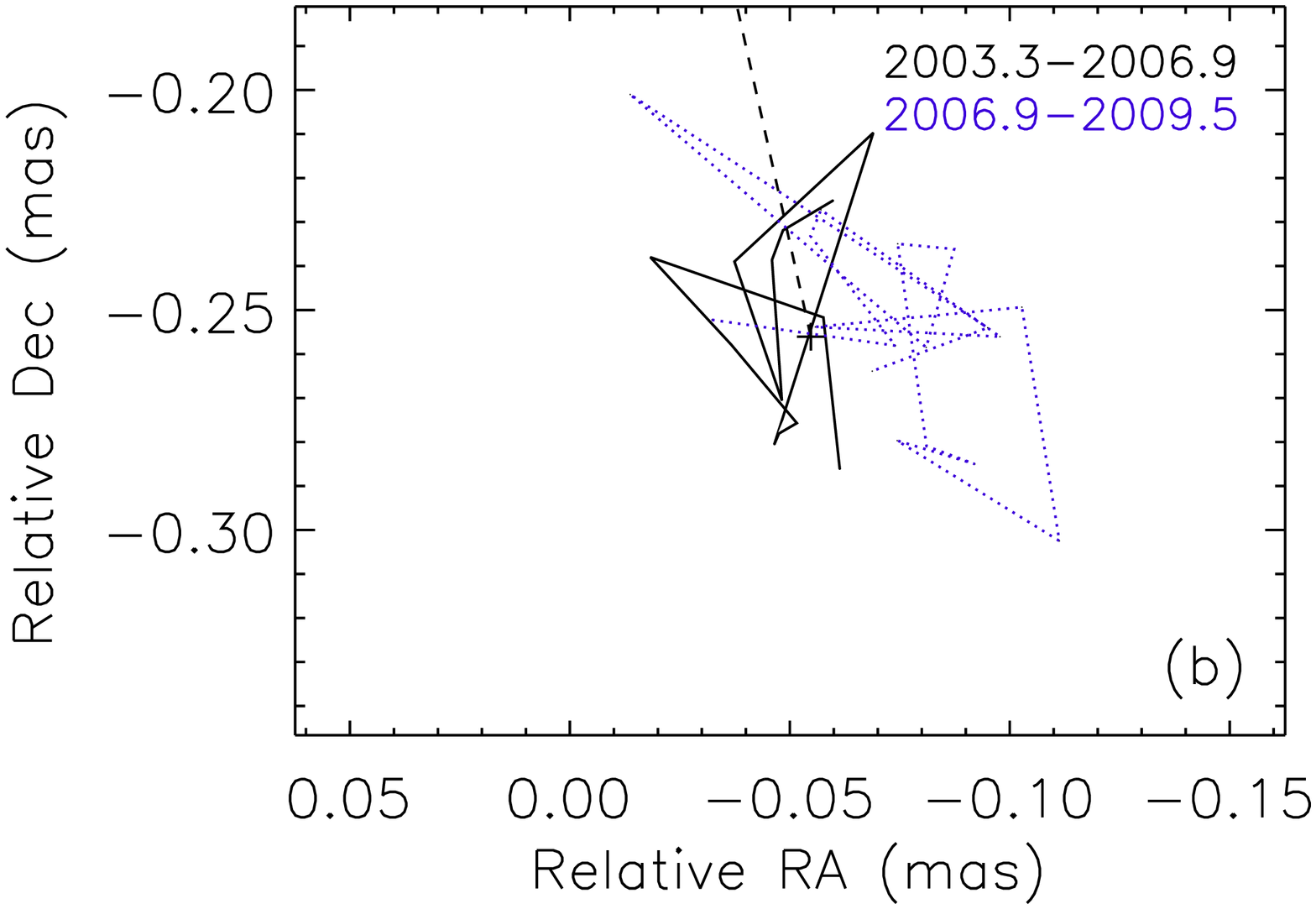}  \\
\includegraphics[width=3.5cm, angle=90]{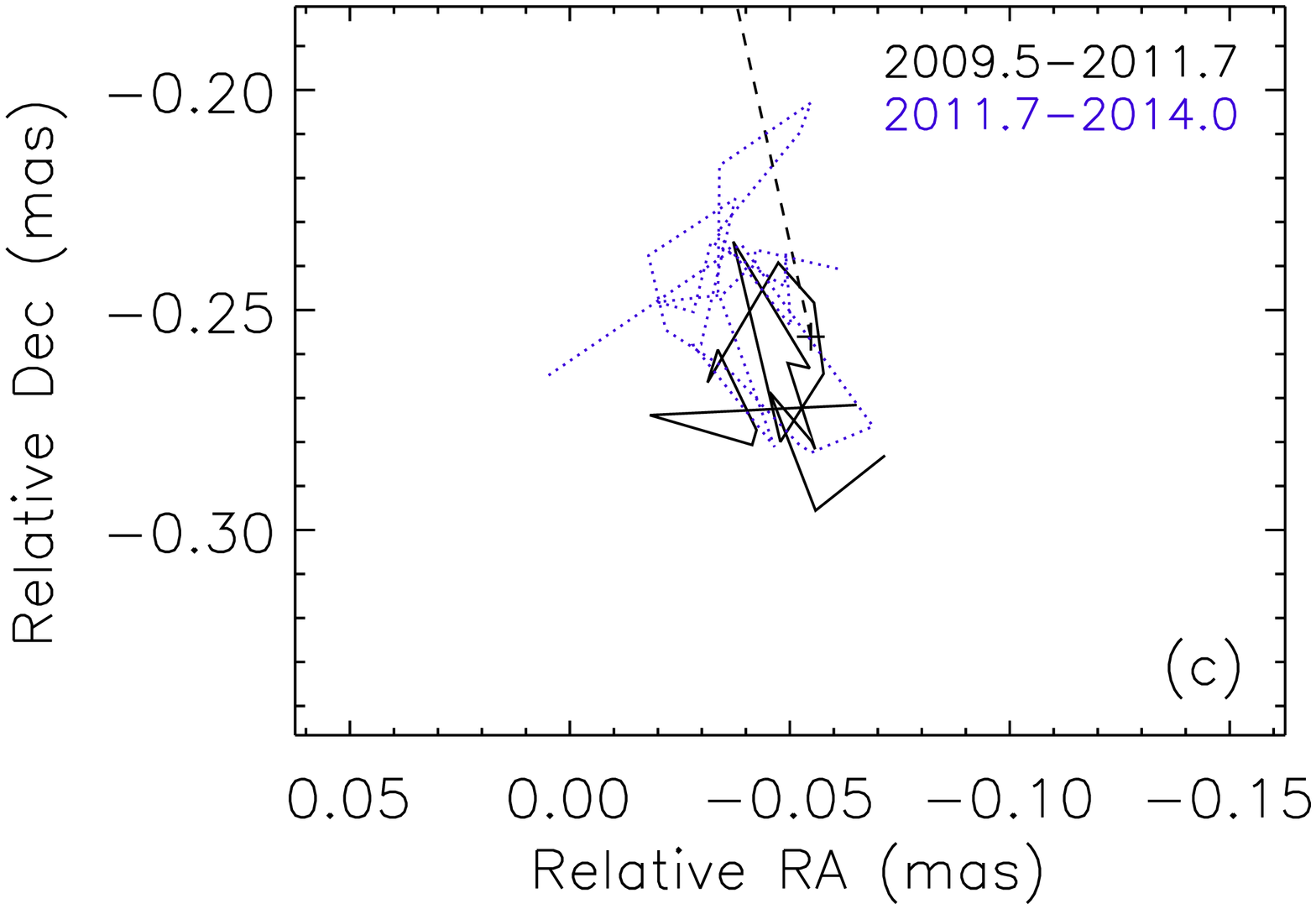}  \\
\includegraphics[width=3.5cm, angle=90]{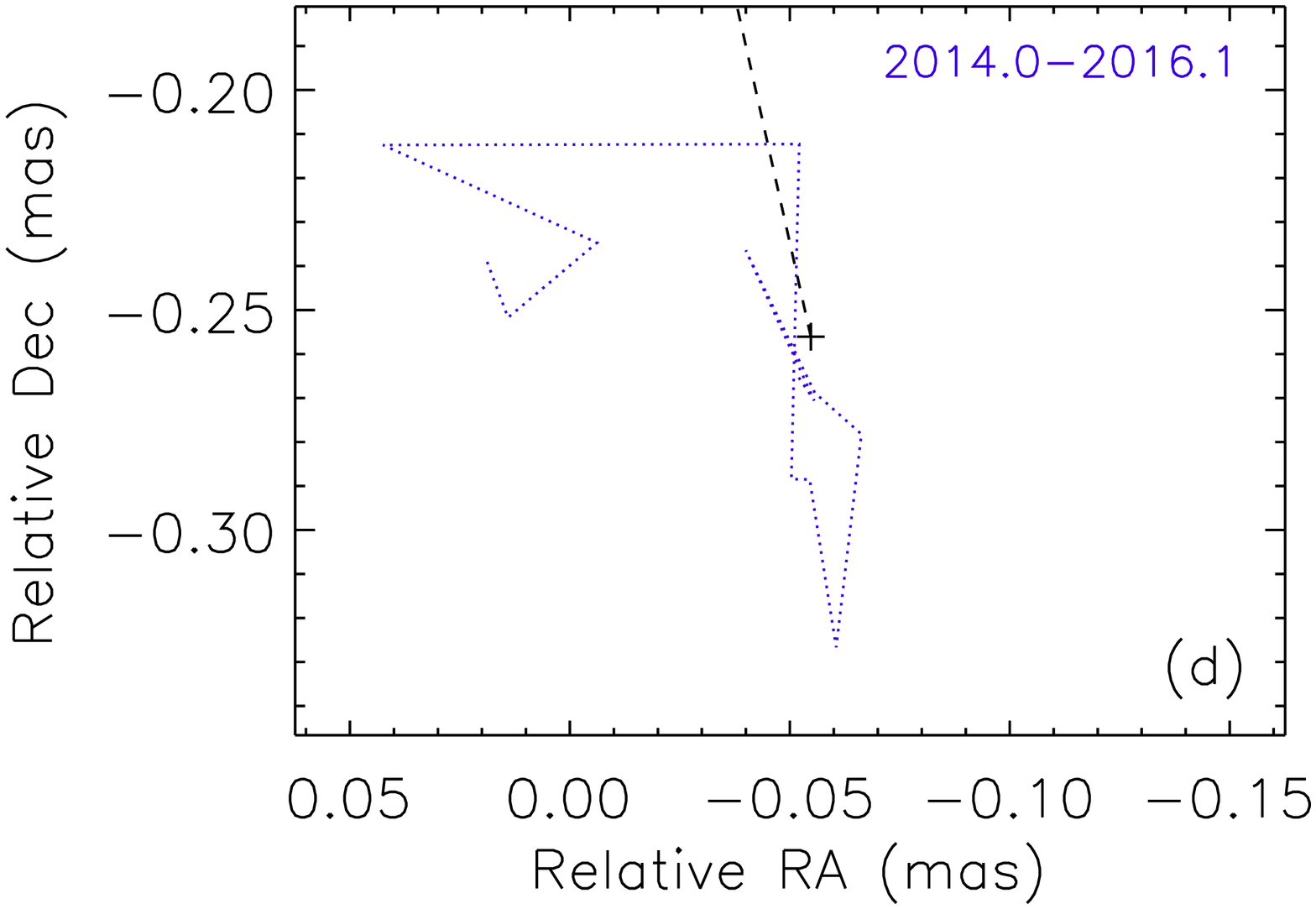}  \\
\includegraphics[width=3.5cm, angle=90]{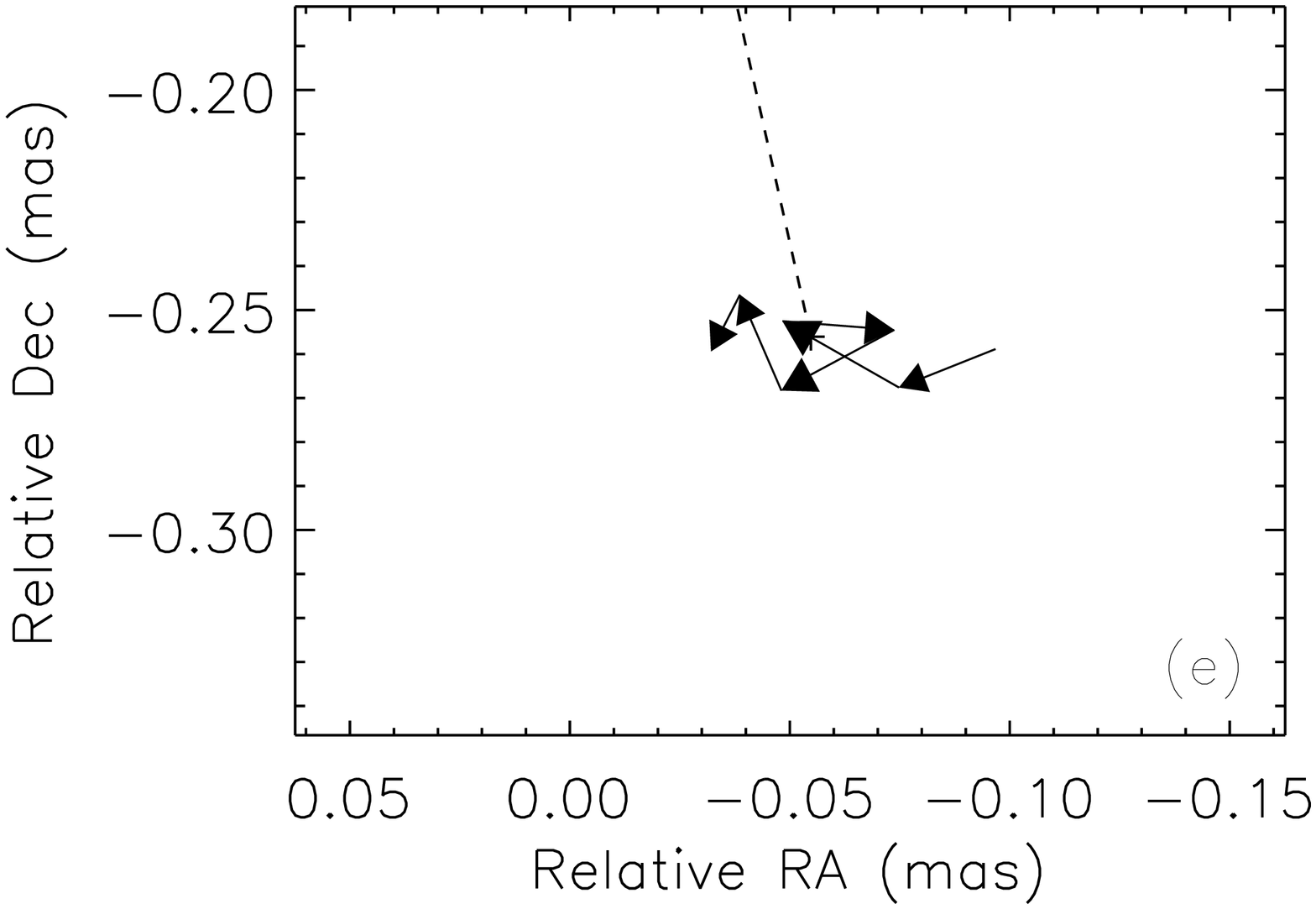}
\caption{Apparent displacements of C7 are shown for seven time periods, 1999.37-2001 and 2001-2003.3 (panel (a), full and dotted blue lines, respectively), 2003.3-2006.9 and 2006.9-2009.5 (b), 2009.5-2011.7 and 2011.7-2014 (c), and 2014-2016.06 (d). Trajectories smoothed over the seven time periods are shown in the panel (e). Arrows indicate the direction of the smoothed motion of C7. The mean error of the lengths of six vectors is 0.008 mas.}
\label{fig:C7_motion}
\end{center}
\end{figure}

\paragraph{Trajectory of motion.}
The best epochs to trace the motion of C7 on time scales of few months are between 2010 and 2013, when the jet is in its stable state and the variance of the  displacements of C7 is therefore at minimum (see Table~\ref{table1}). Variations of the C7 PA are very small and the ridge line of the jet downstream from C7 has a weak variable wiggle \citep{cohen15}. The high cadence of observations during this three year period ($\approx 1.2$ months) allows for a detailed tracking of the C7 trajectory. C7 shows swinging motion with two reversals from clock-wise to anticlockwise and backward to clock-wise on scales of about $0.06$\,pc (Fig.~\ref{fig:C7_motion_2010-2013}, left panel). 

Overall, the  motion of C7 is affected significantly by the core displacements and large observing intervals. To reduce these effects we smooth the apparent motion of C7 in six time intervals. The selection of each time interval comes from consideration of the variation of azimuthal angles of C7 with time (Fig.~\ref{fig:C7_azimuth-epoch}). We define the polar coordinate system with a pole in the median centre of the scatter and polar axis aligned with the jet. We notice that depending upon the epoch, the azimuthal angle is distributed in a specific angular range. It changes within 355-130 degrees for the most of epochs prior to 2003.3, from 20-220 degrees between 2003.3-2006.9, from 150-350 degrees between about 2006.9-2009.5, becomes less than 200 degrees between 2009.5-2011.6 and 2011.6-2014, and becomes smaller beyond 2014. Clustering of azimuthal angles with time suggests that the C7 occupies various regions over 17 years of observation. The apparent displacements of C7 for each time interval are shown in  Fig.~\ref{fig:C7_motion} (a), (b), (c), and (d) panels. Component C7 moves in randomly on time-scales from several days to tens of days. Although the positions of C7 are significantly scattered, there is a tendency for the cluster of positions to move in the clockwise direction. To visualise the smoothed movement of C7, we plot the mean position of C7 estimated over each time interval (Fig.~\ref{fig:C7_motion} (e)), where the arrows indicate the direction of the smoothed motion. 
The first five consecutive vectors show a clockwise rotation with respect to the median position of C7 (in the rest frame of the observer), except the last vector, which is directed anti-clockwise. The direction of the last (sixth) vector is determined by the mean position of C7 estimated over the time interval from 2014 to 2016.06 (Fig.~\ref{fig:C7_motion} (d)). The uncertainty of the mean position is very large (as evident from the large scatter of positions), which makes the direction of the last vector unreliable. The average amplitude of the first five vectors is $0.027\pm0.007$ mas, where the standard errors of each displacement are estimated by propagating the total standard errors of the mean positions of C7 between two consecutive epochs. Excluding the last vector from consideration, we estimate the chance probability that the first five consecutive vectors are oriented in a clockwise direction with respect to the median centre. The probability that a randomly oriented vector has a clockwise direction is $1/2$. The probability of the five random vectors to have a clockwise direction is $(1/2)^{5}=0.031$. The significance of this event to happen by chance is about $3\,\%$ and it is most likely that the C7 performs a clock-wise loop-like motion on time-scales of several years. Regular long-term and high-cadence VLBI observations at 15\,GHz or at a higher frequency of 43\,GHz (such as the ones produced by the Boston University Blazar project\footnote{https://www.bu.edu/blazars/research.html}) are required to identify the trajectory of C7. BL Lac is being also monitored with VLBA at 43\,GHz with a typical cadence of about one month, which is higher to that of the MOJAVE programme.

\begin{figure}[htbp]
\begin{center}
\includegraphics[width=6.2cm, angle=90]{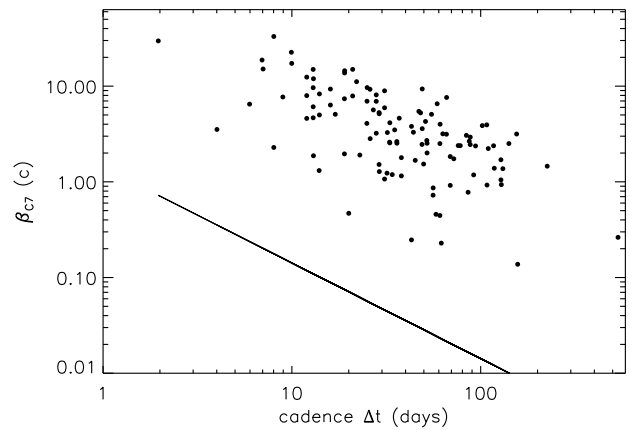}
\caption{Apparent transverse speed of C7 plotted against the time interval between two subsequent observations. $1\sigma$ error bars are presented. The solid line represents the slope of the relation for a maximum apparent displacement, $\Delta r = 2*r_{\rm max} = 0.1$ mas.}
\label{fig:C7_speed-timeint}
\end{center}
\end{figure}

\paragraph{Kinematics.}
We estimate the transverse speeds $\beta_{\rm C7} = \Delta r/(c\,\Delta t)$ at any epoch $t$ using the apparent displacements $\Delta r(t, t+\Delta t)$ passed by C7 during the time $\Delta t$,  where $c$ is the speed of light. The transverse apparent speeds of C7 are mostly superluminal (Fig.~\ref{fig:C7_speed-timeint}) with a mean speed of $4.6c$. Approximately 10\,\% of the speeds are measured to be less than the speed of the light. The apparent speeds are larger on small time scales and remain superluminal on longer time scales. The slope of the relation is due to the fact that the displacement of the C7 is finite and does not depend on the time interval between successive observations. 
To demonstrate it we plot the relation $\beta_{\rm C7} = (2r_{\rm max}/c)\,\Delta t$ (solid line in Fig.~\ref{fig:C7_speed-timeint}), where $2r_{\rm max}=0.1$ mas is the largest possible apparent displacement. The slopes of the observed and simulated relations coincide very well. Superluminal speeds of C7 are hard to explain by a relativistic motion of C7 in a direction close to the line of sight. Cohen et. al. (2014) suggested that the C7 is a RCS, which swings in an irregular manner and excites transverse waves propagating downstream. The jet of the BL Lac is closely aligned to the line of sight and irregular swinging motion of RCS happens in a nearly face-on plane, thus we should expect superluminal speeds in at least 50 \% of motions. Since $\approx 90$ \% of the estimated speeds are superluminal, we assume that the large apparent displacements and, hence, the measured superluminal speeds are most likely due to significant displacements of the core position. To reduce the effect of the core shift, we use smoothed trajectories of the C7 during the seven epochs (Fig.~\ref{fig:C7_motion} (e)). The mean speed is estimated to be subrelativistic, $\langle\beta_{C7}\rangle=0.16\pm0.008$.  The small standard deviation indicates that the mean speed of the C7 remains fairly constant over 15 years.

Let us consider the speeds of C7 during the stable state of the jet. As we discussed earlier in this section, during the stable state the core shift contributes significantly to the apparent motion of C7. To eliminate the contribution of the core we consider projections of the apparent displacements of C7 on the axis transverse to the jet. We calculate the total length of 29 transverse projections (0.24 mas) passed by C7 during three years and estimate the mean transverse speed 1.15$c$, which, in fact, represents the lower limit. Out of 29 measured limiting speeds, 15 are superluminal. This means that most of deprojected speeds on timescales of a month should exceed the speed of light. The large positional errors of C7, which can exceed the apparent displacement between two consecutive positions of C7, may cause unrealistic large transverse projections and, hence, apparent low superluminal speeds.

\begin{figure}
\begin{center}
\includegraphics[width=6.2cm, angle=90]{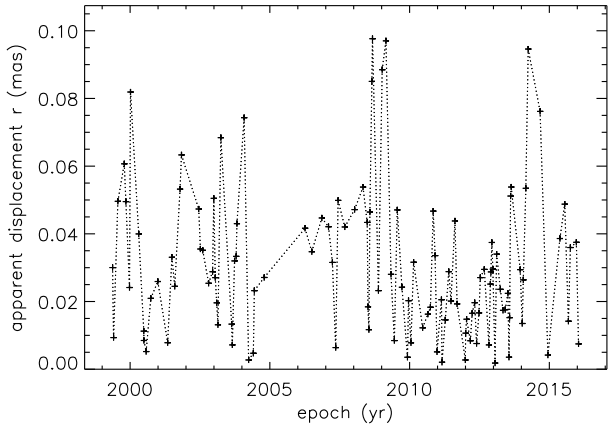}
\caption{Apparent displacements of C7 versus epoch. }
\label{fig:r-epoch}
\end{center}
\end{figure}

\paragraph{Link between the motion of C7 and wave excitation.} The overall jet axis beyond C7 (from 0.26 - 3 mas) has dramatically changed over time \citep{cohen15}. Variation of the position angles of the RCS at 15 GHz, 43 GHz and ridge line of the jet at a distance $\sim 1$ mas from the core also indicate that the PA of the RCS changes on time scales of a few years. \cite{cohen15} showed that the excitation of waves labeled as A, D, and F match with swings in PA of the RCS. The transverse waves excited by shaking of the RCS propagate along the jet with amplitudes ranging from 0.2 mas to 0.9 mas. We then want to check the changes of the apparent displacement of the RCS during the epochs of the wave's excitation.   
The authors concluded that the wave A was excited by RCS between 1998 and 2000.1. From the data available for this period one can see that at epochs $<2000$ the apparent displacements of the RCS are relatively large (Fig.~\ref{fig:r-epoch}). Waves B and C are omitted from consideration since their apparent speeds are not measured because of a poor fit. Wave D with largest amplitude was generated between 2003.5 and 2005. During this time period the RCS has shown an erratic behaviour with maximum apparent displacement $\approx 0.08$ mas, which is very similar to shaking of the RCS. Waves E and F are excited between 2008 and 2009 and we are observing extremely large apparent displacements of the RCS, which reach the maximum apparent displacement $\sim 0.1$ mas. We conclude that the generation of waves A, D, E, and F is accompanied by passing the RCS of relatively large apparent displacements ($> 0.08$ mas) or large amplitudes of shakings of RCS. 

\cite{cohen15} have shown that during the stable jet state (2010-2013) the transverse wave activity becomes less intense, jet ridge lines become quasi-sinusoidal with small and variable amplitudes,  standing features are prominent, the jet PA=$-170.5^\circ$ becomes fairly constant with small wiggle within $\pm 3^{\circ}$. The apparent displacements passed by the RCS and its variance are relatively small (Fig.~\ref{fig:r-epoch}) and trajectory of the RCS shows few reversals. Transverse projections of apparent displacements have a size of $\approx 0.04$ mas and the size of two opposite reversals is about $\approx 0.02$ mas (Fig.~\ref{fig:C7_motion_2010-2013}). It is notable that the size of the transverse projections match with the amplitude of $\approx 0.04$ mas of the transverse wiggling of the PA \citep{cohen15}.     
Moreover, the size of reversals of RCS matches with the maximum amplitude of the ridge line at 2002.94 (Fig.~14 in Cohen et al. 2015). This suggests that during the jet stable state, the RCS acts as the nozzle of the jet and generates quasi-sinusoidal waves with amplitudes lower than $\approx 0.02$ mas.

\section{On-sky brightness asymmetry}
\label{sec:C7_flux_distr} 
\begin{figure}[htbp]
\begin{center}
\includegraphics[width=5.5 cm, angle=-90]{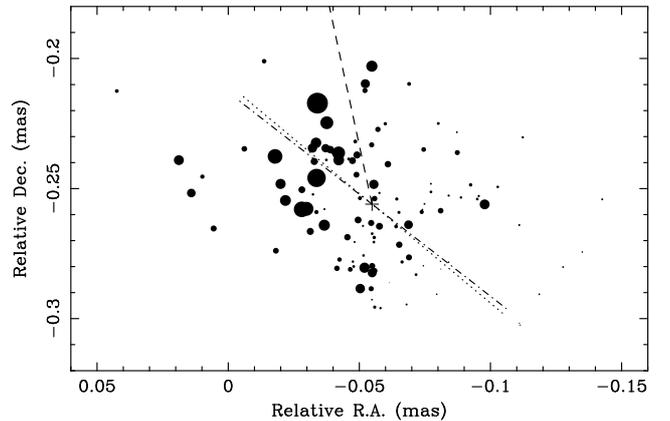} 
\caption{Distribution of flux density measurements of C7 on the sky. The sizes of the circles are proportional to the flux densities of C7, which are in the range from 0.17 Jy to 4.4 Jy. The median position of the scatter is marked by a plus sign. The dashed line connects the median position of C7 and the radio core. The axis of maximum beaming (${\rm PA}_{\rm mb}=-122^{\circ}$) and symmetry axis (${\rm PA}_{\rm sym}=-120^{\circ}$) are denoted by dotted line and dot-dashed line, respectively. }
\label{fig:C7_flux_distr}
\end{center}
\end{figure}

Scatter of 116 positions of C7 on the sky between 1999.37 and 2016.06 is shown in Fig.~\ref{fig:C7_flux_distr}. The sizes of the circles correspond to the flux density of C7 ($f_{\rm C7}$), which ranges from 0.17 Jy to 4.4 Jy.
As can be seen from a visual inspection of Fig.~\ref{fig:C7_flux_distr}, the distribution of flux densities is asymmetric along and transverse to the jet central axis (dashed line). The flux densities tend to weaken down the jet and in a direction transverse to the jet axis. It follows to note that these effects are not result of the core shift since the latter is independent of brightness variation of C7. Any change in the core shift, which may lead to a different degree of asymmetry of displacement vectors, can stretch the observed trajectories of C7 along the jet axis. 

\begin{figure}[htbp]
\begin{center}
\includegraphics[width=6.5cm, angle=90]{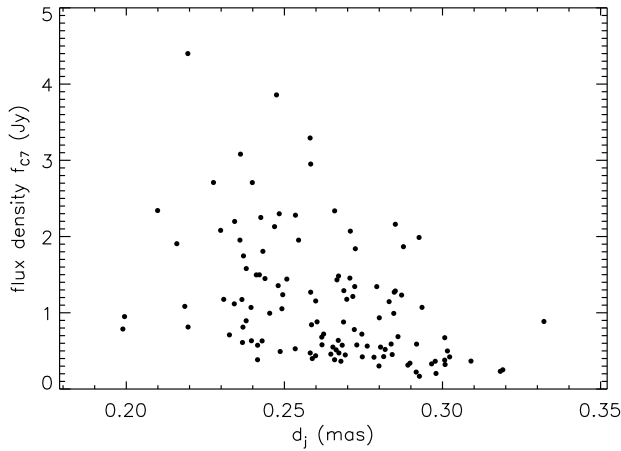}
\caption{Flux density of C7 at 15 GHz ($f_{\rm C7}$) against distance of C7 projected on the jet axis ($d_{\rm j}$). Kendall's rank correlation $\tau=-0.35$ is significant at $p > 0.99$.
}
\label{fig:along_jet}
\end{center}
\end{figure}

To characterise the effect of the brightness asymmetry, we plot the flux density of C7 against its distance from the core projected on the jet axis $d_{\rm j}$ (Fig.~\ref{fig:along_jet}). The spatial distribution of flux densities along the jet axis (Fig.~\ref{fig:along_jet}) shows a tendency for a wider range of flux densities at shorter distances from the core. The flux density range decreases by a factor of four between 0.2 mas and 0.3 mas. Kendall's $\tau$ rank analysis \citep{kendall38} shows that the projected distance of C7 from the core and its flux density are anticorrelated ($\tau=-0.35$) with a high confidence level ($>99.99\,\%$). Brightening of emission upstream in the jet can be explained by a radial velocity field of C7, when the line of sight of the observer is located outside the jet cone. 
\begin{figure}[htbp]
\begin{center}
\includegraphics[width=6.cm, angle=-90]{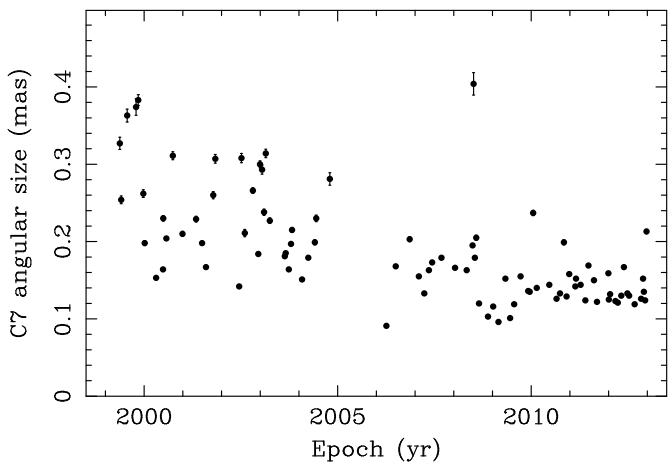}  
\caption{Angular FWHM size of C7 against epoch.}
\label{fig:C7size-time}
\end{center}
\end{figure}
If the steep decrease of the flux density downstream the jet is due to adiabatic cooling of C7, then we should expect an expansion of C7 with time, regardless if it is moving (downstream) or not. The relation between full width at half maximum (FWHM) size of C7 and epoch of observation is shown in Fig.~\ref{fig:C7size-time}. In contrast, the size of C7 on average decreases with time, thus excluding the adiabatic cooling scenario.

\begin{figure}[htbp]
\begin{center}
\includegraphics[width=6.5cm, angle=90]{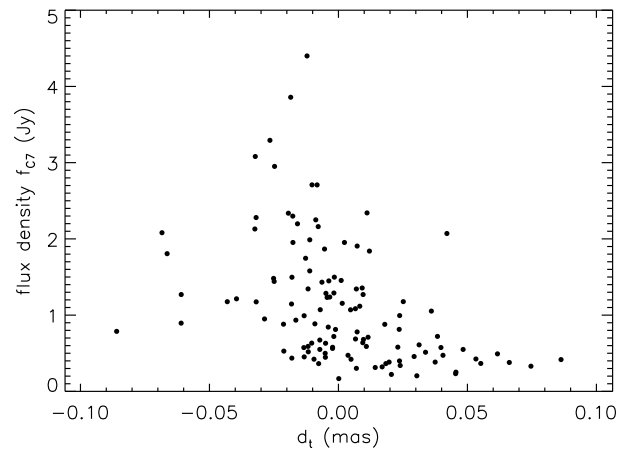}
\caption{Flux density of C7 at 15 GHz ($f_{\rm C7}$) against distance of C7 projected transverse to the jet axis ($d_{\rm t}$). 
Kendall's rank correlation, $\tau=-0.37,$ is significant at $p > 0.99$.
}
\label{fig:trans_jet}
\end{center}
\end{figure}

The emission of C7 is found to be brighter on the east side from the jet axis compared to that of the west (Fig.~\ref{fig:C7_flux_distr}). To characterise this appearance, we define a degree of asymmetry by means of measuring the relationship between $f_{\rm C7}$ and the offset jet distance $d_{\rm t}$ (or projected distances of C7 transverse to the jet axis). The correlation coefficient would be zero for the reflectional symmetric distributions. Kendall's $\tau$ rank analysis  (Fig.~\ref{fig:trans_jet}) shows a significant correlation $\tau=-0.37$ ($p>0.99$) between $f_{\rm C7}$ and $d_{\rm t}$. Thus, the flux density distribution of the C7 is significantly asymmetric with respect to the jet axis. Enhancement of emission can be due to variation of the intrinsic flux of the jet, its speed, and viewing angle or any combination of those. 

Next we estimate the PA at which the emission brightening is the strongest and the flux density distribution has a reflectional symmetry. To determine the PA, we rotate the jet axis by one degree with respect to the median centre of C7 from the ${\rm PA}_{\rm jet}=-167.9^{\circ}$ until ${\rm PA}_{\rm jet}+\pi=12^{\circ}$, and for each given direction, we calculate the projections of the apparent displacements of C7 along that direction ($d_{\rm j}$) and use Kendall's $\tau$ rank method to estimate the correlation coefficient between the projected distances $d_{\rm j}$ of C7 and its flux densities. The position angle ${\rm PA}_{\rm mb}$ at which the correlation coefficient reaches a maximum defines the axis of maximised brightening emission. The maximum correlation coefficient of 0.47 is reached at ${\rm PA}_{\rm mb}=-122^{\circ}$ (dotted line in Fig.~\ref{fig:C7_flux_distr}).  

To estimate the PA of the flux density symmetry axis, we measure the Kenadall's $\tau$ correlation coefficients for each given direction between the transverse distances of C7 and its flux densities and choose the ${\rm PA}_{\rm sym}=-120^{\circ}$ (dot-dashed line in Fig.~\ref{fig:C7_flux_distr}) at which the correlation coefficient is the smallest ($\tau=0.003$). The symmetry axis and maximised beaming axis are aligned remarkably close. The alignment of two independently defined axes means that the beaming axis also can serve as an axis of symmetry, ${\rm PA}_{\rm sym}={\rm PA}_{\rm mb}$. We define the maximised beaming angle $\alpha_{\rm mb}={\rm PA}_{\rm mb} - {\rm PA}_{\rm jet}=47.9^{\circ}$ and symmetry angle $\alpha_{\rm sym}$ between jet axis and symmetry axis, $\alpha_{\rm sym}={\rm PA}_{\rm sym} - {\rm PA}_{\rm jet}=45.9^{\circ}$, which measures the deviation of maximised beaming axis and symmetry axis from the jet direction. 
\begin{figure}
\begin{center}
\includegraphics[width=6.4cm, angle=90]{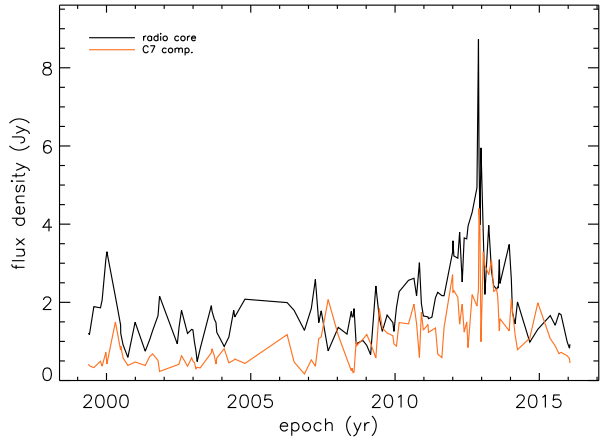}
\caption{Variation of flux density of radio core (black line) and C7 (red line) with time. }
\label{fig:flux_epoch}
\end{center}
\end{figure}

\begin{figure}
\begin{center}
\includegraphics[width=6.2cm, angle=90]{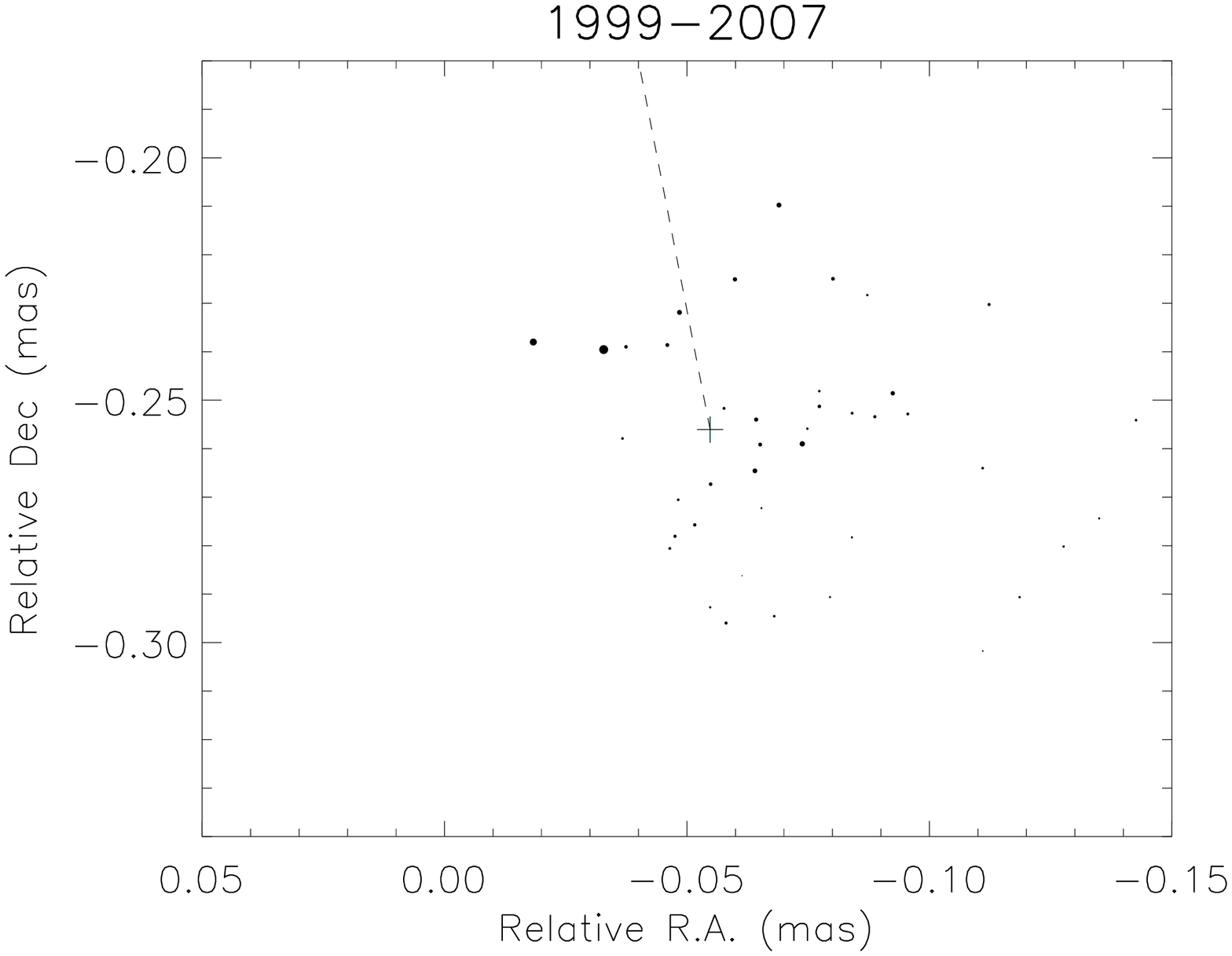} 
\includegraphics[width=6.2cm, angle=90]{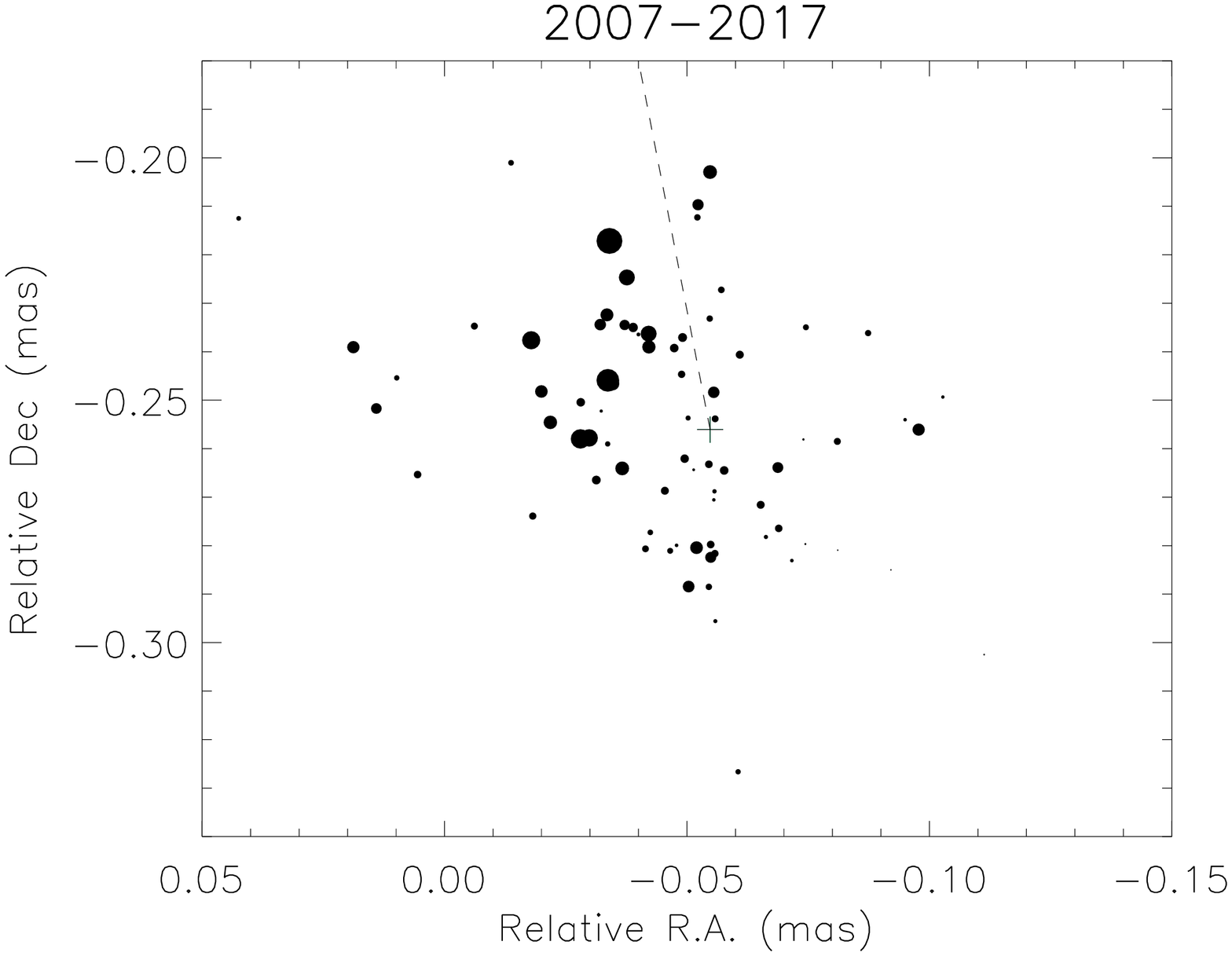} 
\caption{On-sky brightness distribution of C7 during 1999.37-2007 (top panel) and 2007-2016.06 (bottom panel). The sizes of circles are proportional to the values of flux densities from 0.17 Jy to 4.4 Jy. Mean flux densities of C7 are $0.54\pm0.04$ Jy (top panel) and $1.44\pm0.09$ Jy (bottom panel).  }
\label{fig:flux_on-sky}
\end{center}
\end{figure}

\paragraph{Time dependence of brightness asymmetry.}
We examine the change in the brightness asymmetry of C7 over time. For this we divide the sample into two subsamples, one between 1999.37-2007 and another between 2007-2016.06. The choice of the separating epoch 2007 comes from a consideration of the flux density variation (Fig.~\ref{fig:flux_epoch}, red line). It is noticeable that on scales of a few years, the radio emission is relatively low and less variable before 2007, whereas after 2007 there is a radio burst resulting in high flux densities and high variability. The mean flux density at epochs prior to 2007 is about three times lower than that after 2007 (0.54 Jy and 1.44 Jy, respectively). Flux density distributions for the faint subsample (41 epochs between 1999.37-2007) and bright subsample (76 epochs between 2007-2016.06) are shown in Fig.~\ref{fig:flux_on-sky} (top and bottom panels, respectively). The two distributions are positioned asymmetrically with respect to the jet axis: relatively weak flux densities populate the region to the west from the jet direction, while the region to the east is mainly occupied by relatively strong flux densities. This may, in fact, lead to a spurious brightness asymmetry in the direction transverse to the jet if we consider the whole range of epochs from 1999.37 to 2016.06 (Fig.~\ref{fig:C7_flux_distr}), which results in significant negative correlation in the $f_{\rm C7} - d_{\rm t}$ relation plane (Fig.~\ref{fig:trans_jet}). To check it we employ the non-parametric Kendall's $\tau$ test to estimate the correlation coefficient in the $f_{\rm C7} - d_{\rm t}$ relation plane for both subsamples. These are equal to $\tau \approx -0.2$ and have significances of about $95\,\%$ for both subsamples, which indicates that the brightness asymmetry is marginally present in both faint and bright subsamples. In the case of brightening of emission towards the core, that is, the $f_{\rm C7} - d_{\rm j}$ relation plane, the Kendall's $\tau \approx -0.4$ with significance of $>99.9\,\%$ for both subsamples. The brightening of the emission of C7 towards the radio core and transverse to the jet is present in both the weak and the bright subsamples. So, the effect of brightness asymmetry doesn't depend on brightness changes of the jet, which can be caused by variation of the intrinsic flux density and jet speed at the location of C7. Most likely, the brightness asymmetry of C7 is the result of the orientation of the jet at the C7 position to the line of sight. We consider this scenario for a simple velocity field model in the next section.

\begin{figure}
\begin{center}
\includegraphics[width=6.5cm, angle=-90]{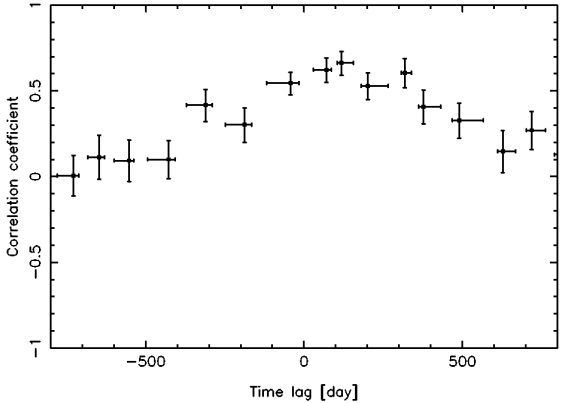}
\caption{Cross-correlation coefficient against time lag for flux density curves of the radio core and C7. }
\label{fig:cc-lag}
\end{center}
\end{figure}

The variation in C7 emission and radio core emission with epoch is shown in Fig.~\ref{fig:flux_epoch}. The utilisation of the z-transformed discrete cross-correlation function tool \citep[zDCF;][]{alexander97} shows what the maximum correlation (0.65) reaches for the time lag of 117 days (0.32 years; Fig.~\ref{fig:cc-lag}). We have to be cautious about artificial correlation due to proximity of the core and C7 - the model-fitting procedure may swap or share the flux between closely spaced features. In Section \ref{sec:pos_errors}, we showed that the flux leakage between the radio core and the C7 is typically 10 \% and, therefore, we believe that the correlation is real. Flux density variation of the core leads that of C7 with a time lag of about four months. The peak of correlation coefficient is blurred, perhaps, due to the core shifts or large dispersion of speeds in the jet plasma. One sigma maximum likelihood intervals of the lag are $+157$ days and $-77$ days, which translates to the 1$\sigma$ lag range from $40$ days to $274$ days. Perturbations in the core move towards C7 with mean apparent speed $0.26 \,\rm{[mas]} / 0.32 \,\rm{[yr]} \approx 0.8$ mas/yr, which translates to the mean apparent speed of $\approx 11c$. The latter is comparable to the measured maximum apparent speed of components $\sim 10\,c$ moving beyond C7 \citep{lister09}. We suggest that the speed of perturbations represents the beam speed of the jet.   

\begin{figure*}[t]
\begin{center}
\includegraphics[width=6.3cm, angle=90]{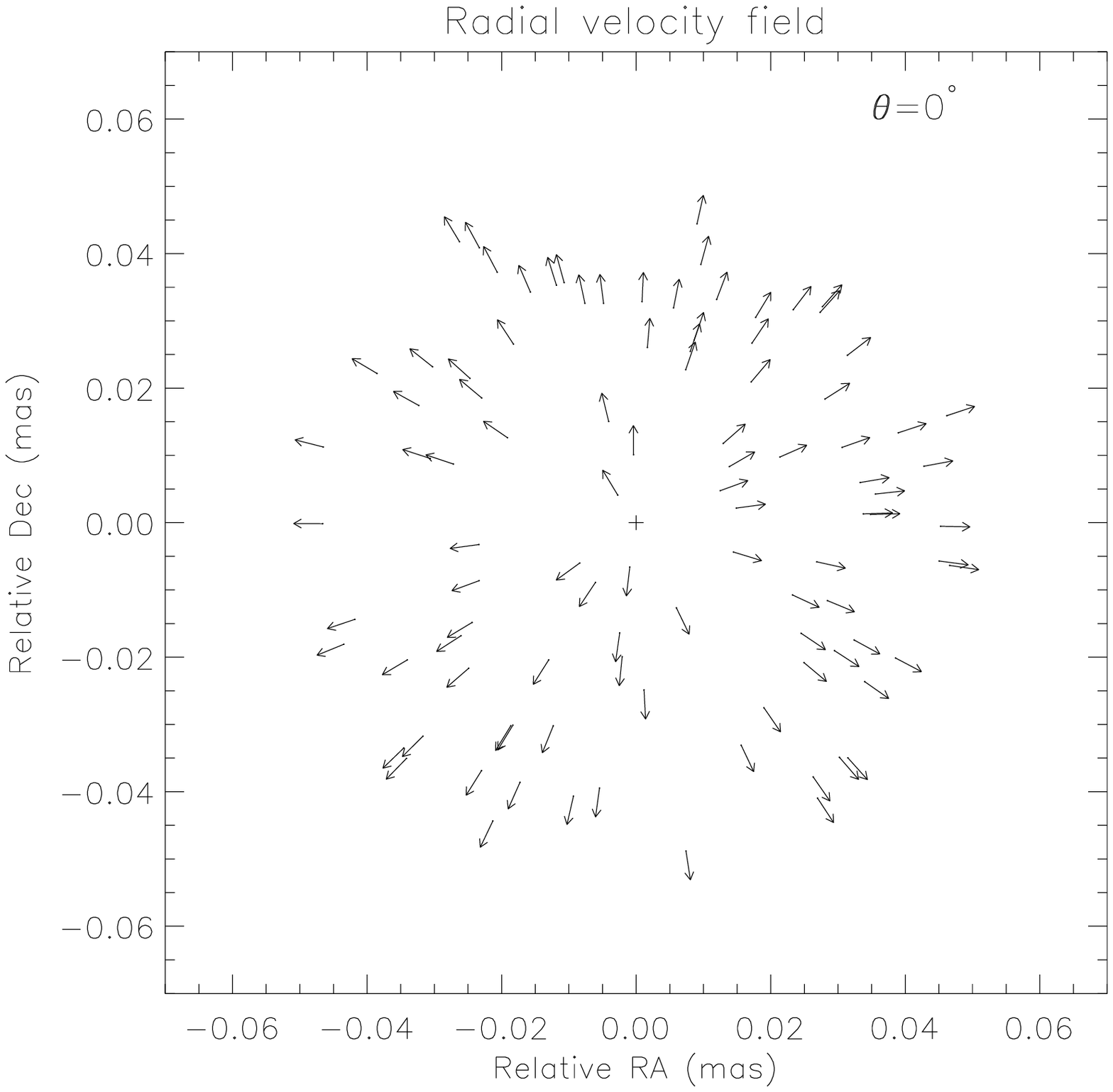}
\includegraphics[width=6.3cm, angle=90]{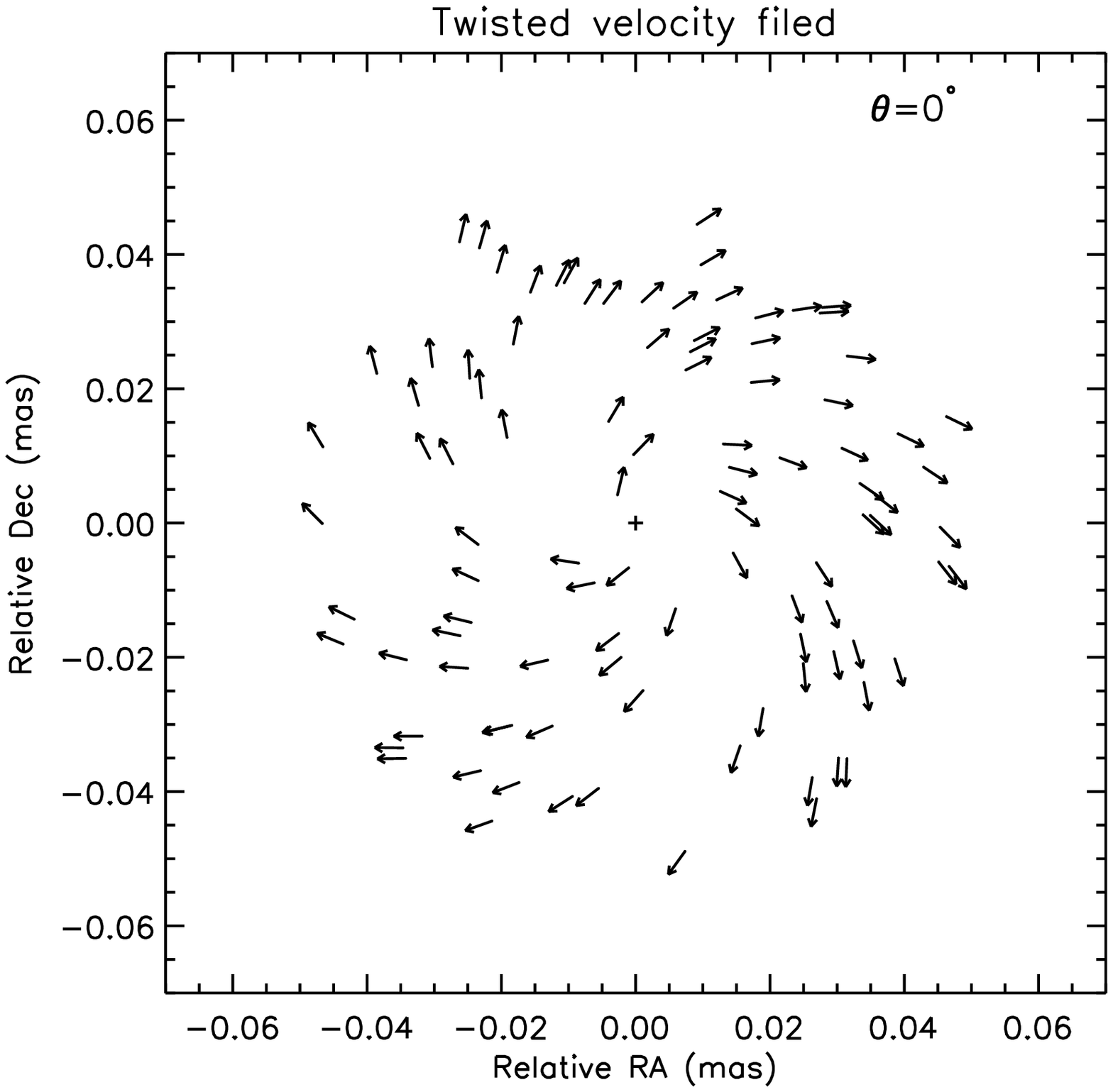}
\caption{On-sky projection of the simulated velocity field with jet viewing angle $\theta = 0^{\circ}$ and Lorentz factor $\gamma=10$. \emph{Left}: Tangential projections of the radial velocity field with $\eta=0^{\circ}$ and $\xi=85^{\circ}$. \emph{Right}: Tangential projections of the twisted velocity field with $\eta=45^{\circ}$ and $\xi=85^{\circ}$. Median position of the scatter of C7 positions is marked by a plus sign. }
\label{fig:rvf}
\end{center}
\end{figure*}

\section{Modelling the brightness asymmetry}
\label{sec:model}

We employ a simple model for the quasi-stationary component C7 in an attempt to identify the configuration of the velocity field of the jet at C7, which is able to reproduce the observed brightness asymmetry. For simplicity, we assume that C7 is the nozzle of the jet, which swings in the plane normal to the jet central axis and drags along the jet stream such that the jet velocity vectors at the nozzle have an axisymmetric distribution with respect to the jet central axis, which, in turn, is inclined at a small angle to the line of sight. The directions of the velocity vectors at the nozzle are limited by a cone with opening angle $\phi$. We assume that the line of the sight is outside of the cone, that is, the jet viewing angle is larger than the half opening angle of the nozzle, $\theta > \phi/2$. We further assume that the intrinsic flux density $f_{\rm int}$ and speed of the jet $\beta c$ are constant and the flux density variability of C7 is due to a change in the orientation of the jet axis with respect to the line of sight. 

The position of the nozzle in the plane of motion is described in polar coordinates, where the pole is located at the median centre of the C7 positions. The polar radius, $r$ (a distance between the pole and the nozzle), and azimuthal angle, $\varphi$ , define the position of the nozzle ($r$, $\varphi$) in the plane of motion. We define the azimuthal angle as positive in the anti-clockwise direction if viewed along the jet central axis. 
\begin{figure*}
\begin{center}
\hspace{0.25cm}\includegraphics[width=6.cm, angle=90]{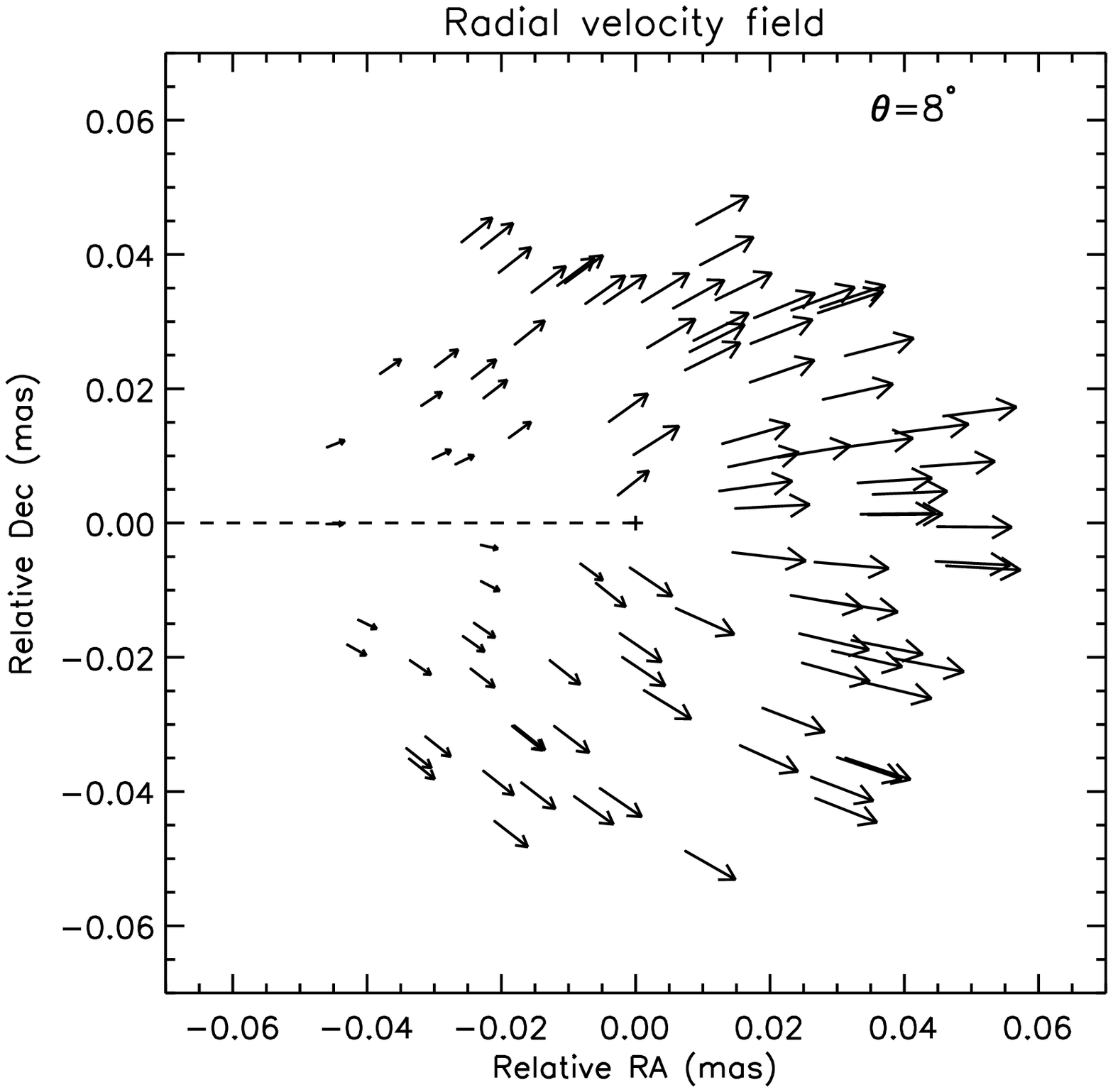}
\hspace{0.38cm}\includegraphics[width=6.cm, angle=90]{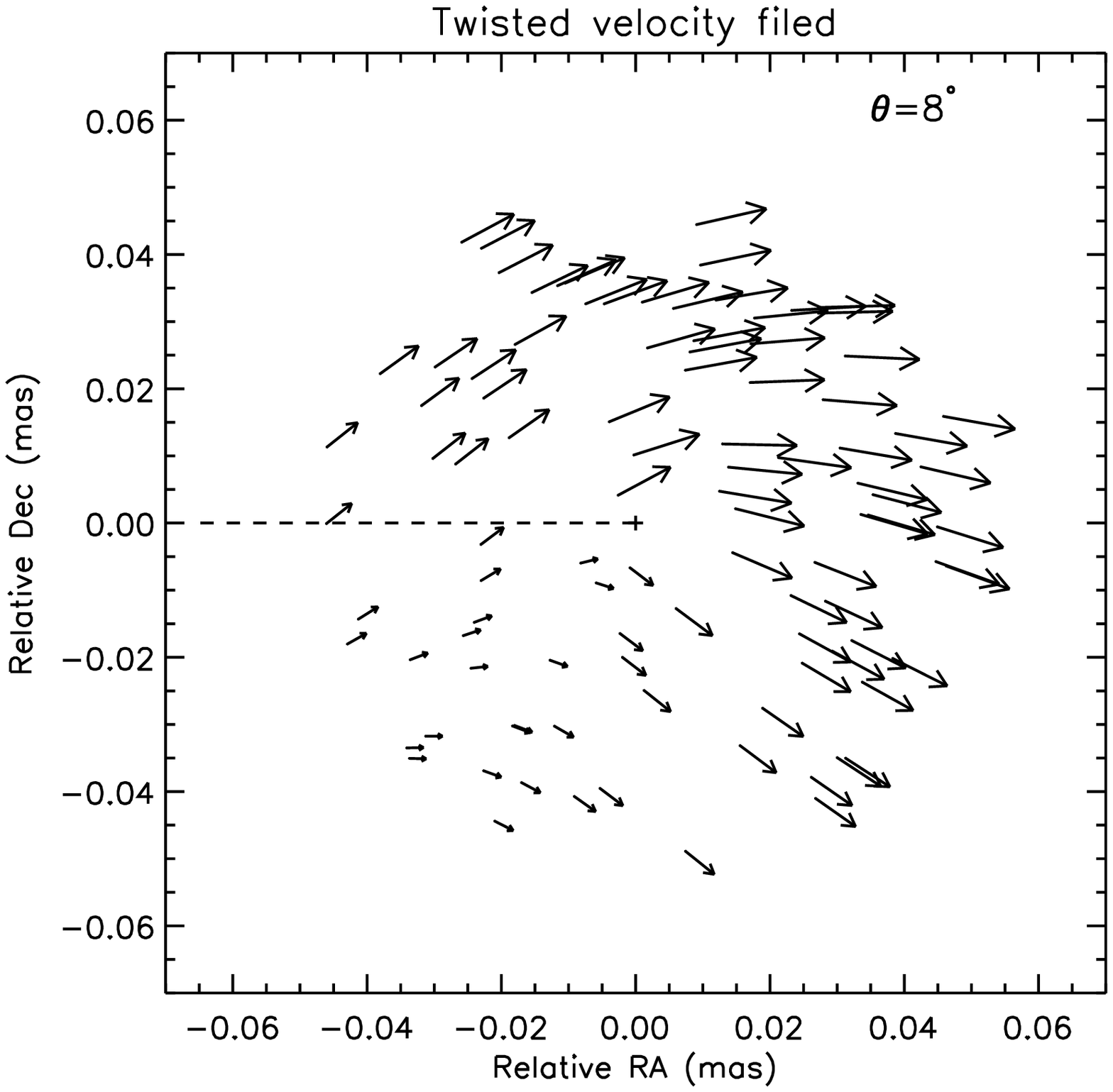}
(a)\includegraphics[width=6.cm, angle=90]{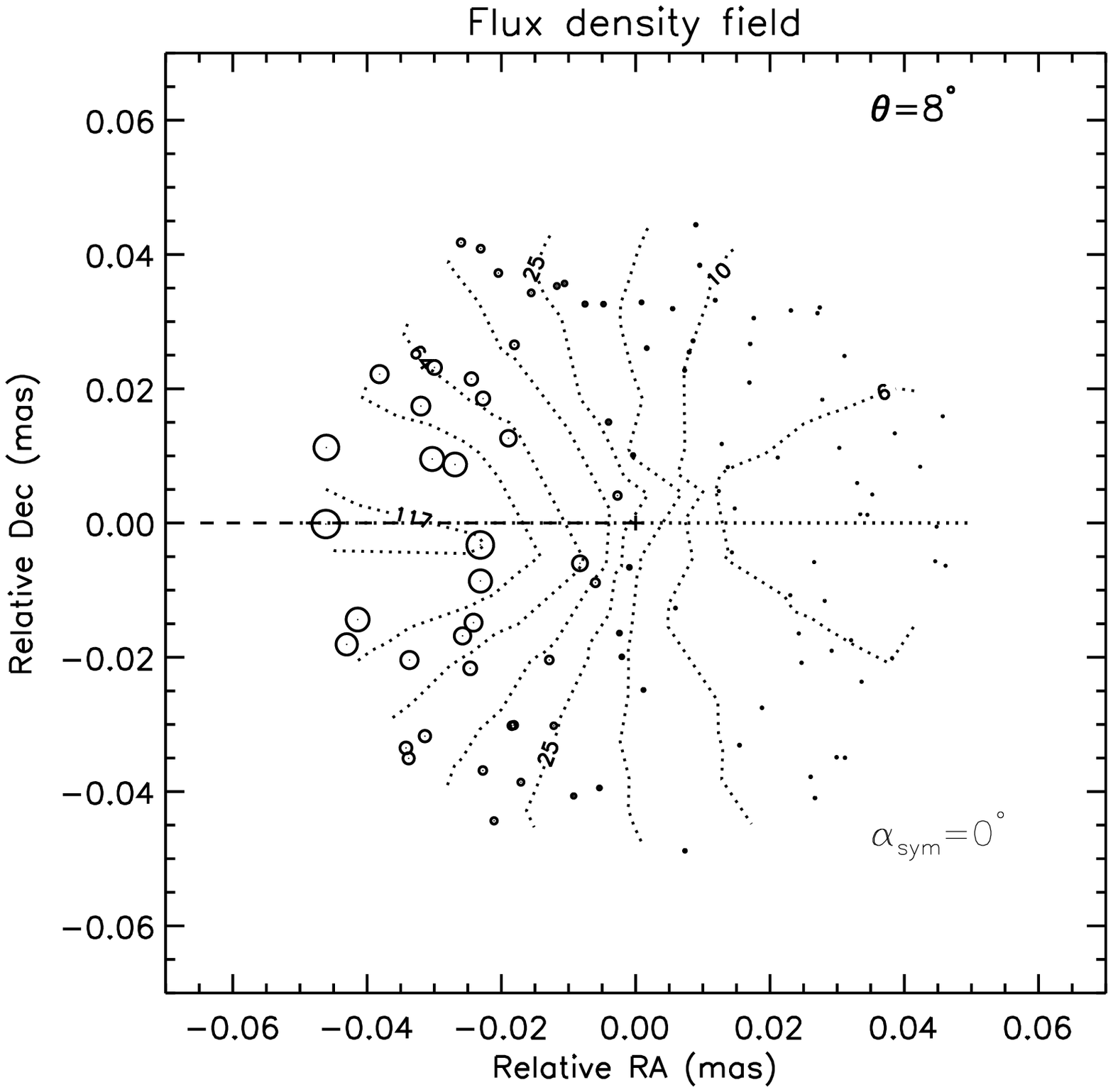}
(b)\includegraphics[width=6.cm, angle=90]{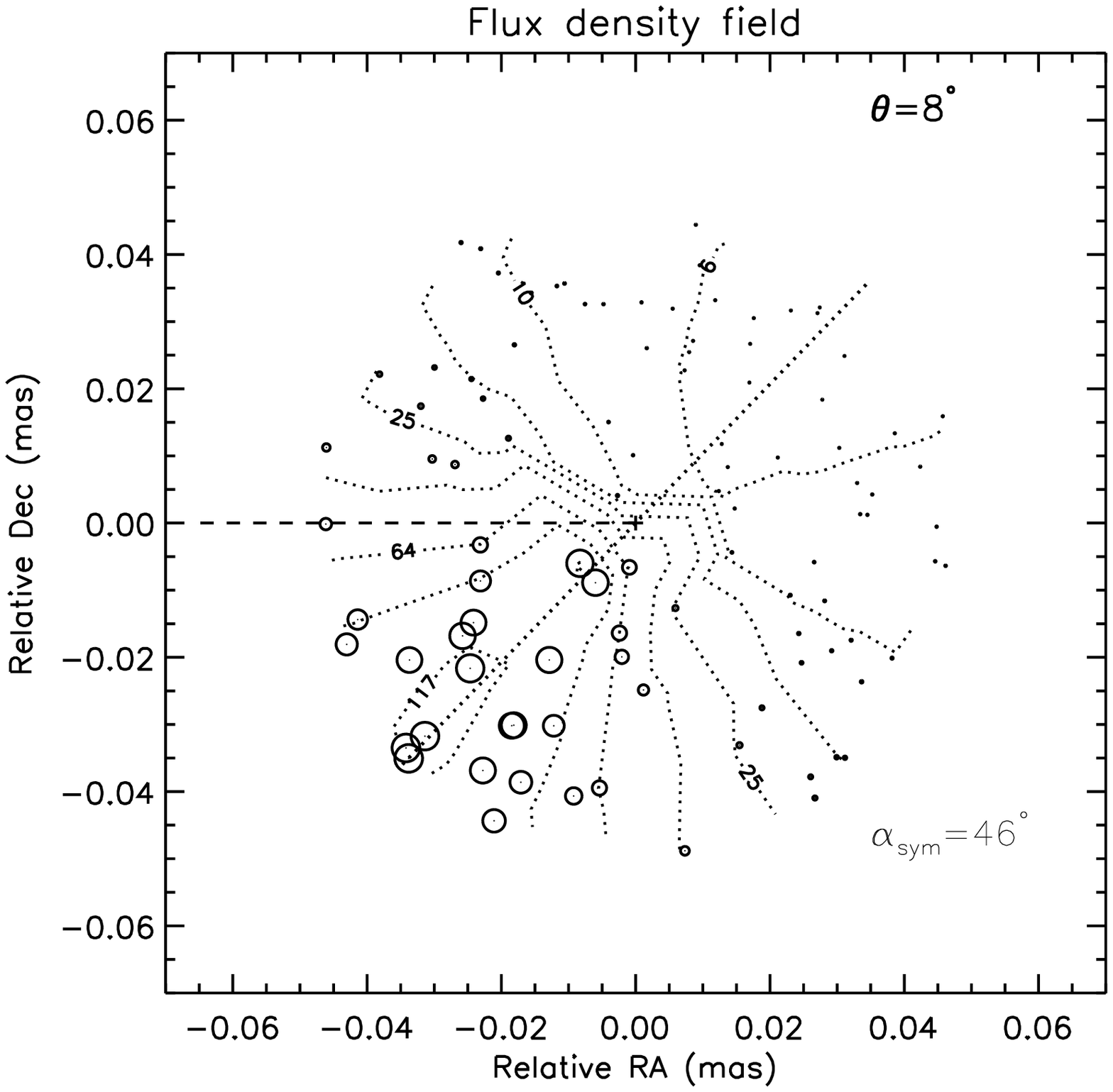}
\caption{On-sky projection of the simulated velocity field and flux density distribution of C7 with the Lorentz factor $\gamma=10$ and jet viewing angle $\theta = 8^{\circ}$ in the case of (a) radial velocity filed (left panels) and (b) twisted velocity field (right panels). Radial and twisted velocity fields are the same as in Fig.~\ref{fig:rvf}. Median position of the scatter of C7 is marked by plus sign. Curved dotted lines represent the contour lines of the flux densities and thick dotted straight line denotes the symmetry axis of the scatter. Projection of the jet axis is shown by dashed line.
}
\label{fig:afd_m1}
\end{center}
\end{figure*}

The direction of the velocity vector of the jet at the nozzle is defined by two angles: $\xi$ is the elevation angle between the velocity vector and the plane of motion and $\eta$ is the azimuthal angle between the projection of a velocity vector on the plane of motion, $r_n = r\cos(\xi)$, and the polar radius. In the case of $\eta=0,$ we recover the radial velocity field with vectors making an angle, $\xi,$ with the polar radius. We note that an angle $\pi/2-\xi = \phi/2$ is the half opening angle of the nozzle. Tangential projection of the radial velocity field on the plane of motion is shown in Fig.~ \ref{fig:rvf} (left panel) for 100 random positions generated within the radius $r_{\rm max}=0.05$ mas.  If $0<\eta<\pi,$ then the velocity field is twisted in a clock-wise direction. The tangential projections of the twisted velocity field are presented in the right panel of Fig.~\ref{fig:rvf} for $\eta = 45^{\circ}$.  

In the relativistic beaming theory the Doppler boosted emission of the jet pattern moving with speed $\beta c$ at the viewing angle $\theta_{\rm C7}$ is given by \citep{pacholczyk70}, 
\begin{equation}
  f_{\rm C7}(r,\varphi) = f_{\rm int}\,D(\beta, \theta_{\rm C7}(r,\varphi))^{p-\alpha},
  \label{eq:sc7}
\end{equation}
where $p = 2$ for a steady-state jet, appropriate for a core region, $\alpha$ is the spectral index assumed to be $\alpha = 0$ for unresolved core components, and the Doppler factor is a function of the jet speed and viewing angle,
\begin{equation}
  D(\beta, \theta_{\rm C7}) = \frac{1}{\gamma(1-\beta\cos(\theta_{\rm C7}(r,\varphi)))},
  \label{eq:dbt}
\end{equation}
and $\gamma=(1-\beta^{2})^{-0.5}$ is the Lorentz factor of the jet flow.

To simulate the `observed' brightness distribution of C7, we generate 100 random positions in the circle of radius $r_{\rm max}=0.05$ mas, which mimic the observed scatter of C7's positions. Each position of C7 is assigned a velocity vector having a constant amplitude $\beta$ and unchanged inclination $\beta(\xi, \eta)$ with respect to the motion plane (or the jet central axis). In turn, the jet central axis makes some angle with the line of sight. The viewing angle of BL Lac's jet was estimated in a number of works discussed in Cohen et al. (2014). Viewing angles are found in the range from $3^{\circ}$ to $12^{\circ}$. The measured maximum apparent speed of the jet $\beta_{\rm app}=9.95  \pm 0.72$ \citep{lister13} puts an upper limit to the jet viewing angle $\theta_{\rm max} \approx 12^{\circ}$ and lower limit to the Lorentz factor of the jet $\gamma_{\rm min} = 10$. We adopt the Lorentz factor of the jet $\gamma = 10$ (or the jet speed $\beta=0.995$) and viewing angle of the jet central axis $\theta = 8^{\circ}$, which results in the apparent speed, $\beta_{\rm app}=7.2$, and the opening angle of the nozzle, $\phi=10^{\circ}$. The latter is chosen so that the half opening angle of the nozzle $\phi/2 = \pi/2 - \xi = 5^{\circ}$ is smaller than the viewing angle of the jet central axis $\theta=8^{\circ}$. We note that the choice of $\phi$ value does not affect our conclusions regarding the simulated brightness asymmetry of C7.

We consider two models of the axisymmetric velocity fields: the radial and twisted velocity fields. The radial velocity field is described by $\eta=0^{\circ}$ and $\xi=90-\phi/2=85^{\circ}$. For the given $\theta = 8^{\circ}$ and intrinsic flux density $f_{\rm int}=0.5$\,Jy, we calculate the viewing angle $\theta_{\rm C7}$ at each position of C7 and, thus, the enhanced flux densities of C7 from Eq.~(\ref{eq:sc7}). The projected velocity vectors of C7 are distributed symmetrically with respect to the jet axis (Fig.~\ref{fig:afd_m1}\,(a), top panel) and their amplitudes become smaller at distances closer to the core. This implies that the line of sight projections of the velocity vectors increase closer to the core and, hence, the Doppler brightening increases upstream (Fig.~\ref{fig:afd_m1}\,(a), bottom panel). The flux density field is symmetric and the symmetry axis is aligned with the jet axis ($\alpha_{\rm sym}\approx0^{\circ}$). The simulated flux density ranges from 3 Jy to 117 Jy.

The direction of a symmetry axis can substantially deviate from the jet axis for non-radial velocity fields, for example, as in the case of the radial velocity vectors, which are twisted at an angle of $\eta = 45^{\circ}$ (Fig.~\ref{fig:afd_m1}(b), top panel). The corresponding flux density field is also symmetric but the axis of symmetry is rotated by $46^{\circ}$ (dotted straight line in Fig.~\ref{fig:afd_m1}\,(b), bottom panel) with respect to the jet axis (dashed line). Brightening of the emission is maximised along the symmetry axis at $\alpha_{\rm sym}=46^{\circ}$ and the moderate brightening exists also upstream. The latter findings mimic those estimated for the observed flux density distribution of C7, that is, $\alpha_{\rm sym} \approx 37^{\circ}$, and the brightening of the C7 emission close to the core (see Fig.~\ref{fig:along_jet}). Thus, the velocity field twisted in a clock-wise direction is able to reproduce the observed brightness asymmetry of C7 and determines the parameter $\alpha_{\rm sym}$, which characterises the direction of the brightness asymmetry. Simulations show that the increase of an azimuthal angle leads to the rotation of the symmetry axis such that $\eta \approx \alpha_{\rm sym}$. The rotation of the symmetry axis by $46^{\circ}$ is shown in the bottom panels of Fig.~\ref{fig:afd_m1}\,(a) and Fig.~\ref{fig:afd_m1}\,(b) for $\eta = 0^{\circ}$ and $\eta=45^{\circ}$, respectively.  

For the given jet viewing angle, $\theta=8^{\circ}$, the radial velocity field of the nozzle can produce the flux density brightening upstream but not the observed brightness asymmetry. The latter can be modelled with the twisted axisymmetric velocity field, which forms by a swinging of the jet nozzle. A more sophisticated model of the velocity field is needed to fully describe the characteristics of the observed brightness asymmetry. We will elaborate on this in an upcoming paper.

\section{Discussion}
\label{sec:discussions}

In Section \ref{sec:motion}, we found that the  displacements of the core in the jet direction can dominate over the  displacements of C7. We considered what the possible origins of the core shifts . One of the plausible scenarios is the resolution-dependent core shift. \cite{cohen14} noted that the core of BL Lac at 15 GHz is a compound core, which consists of two stationary features observed with high-resolution VLBA at 43 GHz \citep{jorstad05}. The beam size at 15 GHz is $\sim 1$ mas, which is not sufficient for  resolving these two features positioned at 0.1 mas angular displacement. The relative brightness of these two stationary features results in movement of the 15 GHz core in the direction of the jet axis. The core shift may happen because of physical and geometric effects, that is, the change of opacity during the radio flares and wobbling of the jet. The opacity change is due to an increase of particle density or amplitude of the magnetic field. Wobbling of the jet can be produced either by variations of flow injection, which may result from changes in the particle density or magnetic field configuration caused by either turbulence or irregularities in the accretion process \citep[e.g.][]{agudo12}, or nuclear flares, which significantly increase the density of emitting relativistic particles in the core region \citep{plavin19}, or from orbital motion of compact objects in a binary black hole system \citep{valtonen06}. Projection effects can also cause the core shifts in the case of the large apparent PA variations of the jet and sharp jet bends if the jet viewing angle is small. 

It should be noted that only the resolution-dependent core shift and change of opacity are capable of producing the core shifts along the jet axis. In other scenarios, the core shift may happen in random directions depending on the positioning of the jet with respect to the line of sight. 

For BL Lac, the core shift contributes significantly to the apparent motion of C7 (or RCS). The significant contribution of the core shift and probable large positional errors of C7 leads to blurs and makes it impossible to precisely trace the  motion of C7. Ideal candidates among blazars for studying the  motion of stationary components are those with an insignificant core shift. A high-cadence VLBA monitoring at high radio frequencies (e.g. 43 GHz) would be advantageous for escaping resolution-dependent core shift effects, tracking the dynamics, and testing the rotation of C7.   

During the jet stability period, the brightness of the core and RCS increases erratically between 2011.5 and 2013.5 (red line in Fig.~\ref{fig:flux_epoch}). Such unusually strong activity is also observed in a wide electromagnetic spectrum from millimetre through gamma-rays between 2011.4 and 2012.8 \citep{reiteri13}. Variation of the speed and viewing angle of the jet cannot cause strong flux density variability since the pattern speed of moving components do not show an increase between 2010 and 2013 \citep{cohen14} and the PA of the RCS (or the viewing angle of the jet) remain almost unchanged \citep{cohen15}. Interplay between the jet plasma flow (or accretion rate activity) and the strength of a helical magnetic field of the jet can be responsible for the exceptional flare activity. It remains unclear if the state of the jet stability and the late event of the violent radio flux variability are physically related or if they are two independent superimposed events.

\section{Conclusions}
\label{sec:summary}

The principal results of our analysis of 116 positions of the quasi-stationary component C7 of the jet in BL Lac can be summarised as follows:
\begin{itemize}
 
 \item The motion of C7 is limited to an area of about $0.1$\,mas ($0.13$\,pc). The estimated positional errors of C7 represent the lower limits. The average value of the position errors along the jet axis ($7.6$ $\mu$as) is larger than that in the transverse direction ($2.1$ $\mu$as) by a factor of 3.5. The typical upper limits of positional errors are larger than a few times the lower limits. Simulations show that the proximity of the bright core and C7 does not lead to a spurious dependence between the C7 position and its flux and that the flux leakage between C7 and the radio core is typically within $10\,\%$.
 
 \item The apparent displacement vectors of C7 have an anisotropic distribution and show asymmetry in the jet direction with tendency to be longer along the jet axis. These effects are most likely due to resolution-dependent core shift or opacity effects. 
 
 \item We developed a statistical tool for estimating the mean and variance of the displacements of the core and C7. We find that the  displacements of C7 (isotropic component) and the core (anisotropic component) have an almost commensurate contribution to the apparent displacements of C7. The estimates of statistical characteristics of displacements are close to true values for short observation intervals ($<35$ days). The rms of displacements of the core and C7 decreases 1.5-2 times during the stable jet state compared to those during the swinging activity of C7. The contribution of the core and C7 to the apparent motion of C7 remains of comparable importance during both states of activity. The rms of the spatial displacements of the core is a factor of four larger than those of the C7. 
 
 \item During the jet stability period, the motion of the C7 on time scales of months is complex, showing swinging motion with reversals of the direction, and superluminal speeds. The trajectories of C7, smoothed over time scales of a few years, show a clock-wise loop motion with mean subrelativistic speed $(0.16\pm0.008)\,c$.
 
 \item Long-term monitoring with high cadence (less than or about a month) VLBA observations at 15 GHz is needed to study the dynamics of the C7. Accurate and low positional uncertainties of the C7 are crucial for identifying the dominant contribution of the intrinsic motions of the core and C7 to the apparent motion of C7.
 
 \item We confirm that the excitation of transverse waves moving along the jet is generated by the motion of the C7, which acts as the nozzle to the jet. The excitation of transverse waves with relatively large amplitudes ($\gtrsim 0.2$ mas) are associated with excessively large apparent displacement vectors ($>0.08$ mas) of the nozzle. Quasi-sinusoidal stable waves with small amplitudes ($\lesssim 0.02$ mas) are generated by the reversal motion of the nozzle within the angular distance of $\approx 0.02$ mas ($0.03$\,pc), which matches nicely with the amplitudes of the excited waves.
 
 \item The on-sky distribution of flux densities of C7 reveals a statistically significant brightening of the emission upstream and transverse to the jet. The observed brightness asymmetry is not dependent on the epoch and it is, rather, associated with a regular change of the jet viewing angle at the location of the C7 (or the nozzle).
 
 \item The asymmetry of the brightness can be reproduced by a toy model of C7. The model assumes an axisymmetric twisted jet velocity field for the C7 and a jet viewing angle of $8^{\circ}$, while the intrinsic flux density and speed of the jet are unchanged. More sophisticated models are necessary to recover the jet viewing angle of the BL Lac and parameters characterising the velocity field of the jet.

\end{itemize}

\begin{acknowledgements}
We thank Eduardo Ros for a careful reading the manuscript and valuable comments and Marshall Cohen for useful discussions and comments. A.B.P. was supported by the Russian Science Foundation grant 16-12-10481. T.S. was supported by the Academy of Finland projects 274477, 284495 and 312496. The VLBA is a facility of the National Radio Astronomy Observatory, a facility of the National Science Foundation that is operated under cooperative agreement with Associated Universities, Inc. This research has made use of data from the MOJAVE database that is maintained by the MOJAVE team \citep{lister18}. 
\end{acknowledgements}

\begin{appendix}
\section{Standard error of the rms}
\label{asec:se_rms}
We assume that the measurement errors $\delta_{r_n}$ and $\delta_{r_j}$ of the displacements, projected transverse to the jet axis and along the jet axis ($r_n$ and $r_j$), have Gaussian distributions.  

The rms${_s}$ of displacements of C7 is a function of $r_n$ (see Eq.~(\ref{eq:rms_s})). 
Our task is to derive the standard error of the rms${_s}$ ($\delta_{{\rm rms}_{s}}$) given the measurement errors 
$\delta_{r_n}$. We apply the basic formula for error propagation to Eq.~(\ref{eq:rms_s}),
\begin{equation}
        \delta_{{\rm rms}_{s}} = \frac{\delta_{\overline{r_n^{2}}}} {\sqrt{2 \overline{r_n^{2}}}}.
        \label{aeq:se_rms_s}
\end{equation}
To express $\delta_{\overline{r_n^{2}}}$ trough $\delta_{r_n}$, first, we use the formula for power to derive $\delta_{r_n^{2}} = 2r_n \delta_{r_n}$. Next we propagate the standard error of $\overline{r_n^{2}}$, which, according to variance error propagation, can be written as,
\begin{equation}
        \delta_{\overline{r_n^{2}}} = \frac{\sqrt{\sum{\delta^2_{r_n}}}} {N}  = \sqrt{ \frac{\overline{\delta^2_{r_n}}} {N} },
        \label{aeq:se_mrn2}
\end{equation}
where $N$ is the number of displacements. Lastly, substituting the $\delta_{\overline{r_n^{2}}}$ into Eq.~(\ref{aeq:se_rms_s}), we obtain,
\begin{equation}
        \delta_{{\rm rms}_{s}} = \sqrt{ \frac{ \overline{\delta^2_{r_n}} } { 2 \overline{r_n^{2}}} }.
        \label{aeq:se_rmss}
\end{equation}
\\

The rms${_c}$ of the core displacements is a complex function of $r_n$ and $r_j$ (see Eq.~(\ref{eq:rms_c})).  
To derive the standard error of rms$_c$, we use a step by step approach to carefully propagate an error in Eq.~(\ref{eq:rms_c}). First, we derive the standard errors of $\delta^2_{r_j}$ and $\delta^2_{r_n}$ and use the error propagation rules to derive $\delta_{{\rm rms}_{c}}$. 

Let us represent the standard deviation of ${r_j}$ in the form, $\sigma^2_{r_j} = \overline{r_j^2} - \overline{r_j}^2$.
Our first task is to propagate the errors for $\overline{r_j^2}$ and $\overline{r_j}^2$. The error in the $\overline{r_j}$ is given by $\delta_{\overline{r_j}} = \overline{(\delta^2_{r_j}}/N)^{0.5}$. Then using the formula for powers we obtain the error of $\overline{r_j}^2$,
\begin{equation}
        \delta_{\overline{r_j}^2} = 2 \overline{r_j}^2 \delta_{\overline{r_j}}.
        \label{aeq:se_m2rj}
\end{equation}

We derive the uncertainty of $\overline{r_j}^2$ in two steps. First, we propagate an error in $r_j^2$, $\delta_{r_j^2} = 2 r_j^2 \delta_{r_j}$, and then derive the error of $\overline{r_j^2}$ (see Eq.~(\ref{aeq:se_mrn2})),
\begin{equation}
        \delta_{\overline{r_j^2}} = \sqrt{ \frac{\overline{\delta^2_{r_j^2}}} {N}  }.
        \label{aeq:se_mrj2}
\end{equation}
Having the standard errors of $\overline{r_j}^2$ and $\overline{r_j^2}$ (Eqs.~(\ref{aeq:se_m2rj}) and (\ref{aeq:se_mrj2})), we obtain the error of their difference, $\sigma^2_{r_j} = \overline{r_j^2} - \overline{r_j}^2$,
\begin{equation}
        \delta_{\sigma^2_{r_j}} = \sqrt{ \delta^2_{\overline{r_j^2}} + \delta^2_{\overline{r_j}^2} } = \sqrt{ \frac{\overline{\delta^2_{r_j^2}}} {N} +  (2 \overline{r_j}^2 \delta_{\overline{r_j}})^2 }.   
        \label{aeq:se_se2_rj}
\end{equation}
Similarly, we derive the error for the ${\sigma^2_{r_n}}$, 
\begin{equation}
        \delta_{\sigma^2_{r_n}} = \sqrt{ \frac{\overline{\delta^2_{r_n^2}}} {N} +  (2 \overline{r_n}^2 \delta_{\overline{r_n}})^2 }. 
        \label{aeq:se_se2_rn}
\end{equation}
The error of the difference $\sigma^2_{r_j} - \sigma^2_{r_n} = {\rm rms}^2_c$ is then given by,
\begin{equation}
        \delta_{{\rm rms}^2_c} = \sqrt{\delta^2_{\sigma^2_{r_j}} + \delta^2_{\sigma^2_{r_n}} },      \label{aeq:se_rms2c}
\end{equation}
and, finally, the error of the rms$_c = ( \sigma^2_{r_j} - \sigma^2_{r_n} )^{0.5}$ is derived using a power error propagation:

\begin{equation}
        \delta_{{\rm rms}_c} = \frac{{\rm rms}_c \, \delta_{{\rm rms}^2_c} } {2 ( \sigma^2_{r_j} - \sigma^2_{r_n} )}.
        \label{aeq:se_rmsc}
\end{equation}
The errors of the rms for spatial displacements of the C7 and core are:
\begin{equation}
        \delta_{{\rm rms}_{S}} = \sqrt{\frac{3}{2}} \,\delta_{{\rm rms}_{s}} \approx 1.3 \, \delta_{{\rm rms}_{s}}
        \label{aeq:se_rms_S}
\end{equation}
and
\begin{equation}
        \delta_{{\rm rms}_{C}} = \frac{\delta_{{\rm rms}_{c}}} {\sin(\theta)} \approx 7.2 \,\delta_{{\rm rms}_{c}}
        \label{aeq:se_rms_C}
\end{equation}
for $\theta = 8^{\circ}$.

\end{appendix}

\bibliographystyle{aa}
\bibliography{refs_tga}

\end{document}